\documentclass[twocolumn,prb,amsmath,showpacs,amssymb]{revtex4}
\usepackage{graphicx}
\usepackage{dcolumn}
\usepackage{bm}

\begin{document}


\title{Magnetic dipolar ordering and relaxation in the high-spin molecular cluster compound Mn$_{6}$}
\author{A. Morello$^*$}
\affiliation{Kamerlingh Onnes Laboratory, Leiden Institute of
Physics, Leiden University, P.O. Box 9504, 2300 RA
 Leiden, The Netherlands}
\author{F. L. Mettes}
\affiliation{Kamerlingh Onnes Laboratory, Leiden Institute of
Physics, Leiden University, P.O. Box 9504, 2300 RA
 Leiden, The Netherlands}
\author{O. N. Bakharev}
\affiliation{Kamerlingh Onnes Laboratory, Leiden Institute of
Physics, Leiden University, P.O. Box 9504, 2300 RA
 Leiden, The Netherlands}
\author{H. B. Brom}
\affiliation{Kamerlingh Onnes Laboratory, Leiden Institute of
Physics, Leiden University, P.O. Box 9504, 2300 RA
 Leiden, The Netherlands}
\author{L. J. de Jongh}
\affiliation{Kamerlingh Onnes Laboratory, Leiden Institute of
Physics, Leiden University, P.O. Box 9504, 2300 RA
 Leiden, The Netherlands}
\author{F. Luis}
\affiliation{Instituto de Ciencia de Materiales de Arag\'on,
CSIC-Universidad de Zaragoza, 50009 Zaragoza, Spain.}
\author{J. F. Fern\'{a}ndez}
\affiliation{Instituto de Ciencia de Materiales de Arag\'on,
CSIC-Universidad de Zaragoza, 50009 Zaragoza, Spain.}
\author{G. Arom\'{\i}}
\affiliation{Departament de Qu\'{\i}mica Inorg\'{a}nica,
Universitat de Barcelona, Av. Diagonal 647, 08028 Barcelona,
Spain.}


\date{\today}

\begin{abstract}
Few examples of magnetic systems displaying a transition to pure
dipolar magnetic order are known to date. As was recently shown,
within the newly discovered class of single molecule magnets quite
attractive examples of dipolar magnetism may be found. The
molecular cluster spins and thus their dipolar interaction energy
can be quite high, leading to reasonably accessible ordering
temperatures even for sizable intercluster distances. In favorable
cases bonding between clusters in the molecular crystal is by Van
der Waals forces only, and no exchange paths of importance can be
distinguished. An important restriction, however, is the
requirement of sufficiently  low crystal field anisotropy for the
cluster spin, in order to prevent the occurrence of
superparamagnetic blocking at temperatures above the dipolar
ordering transition. This condition can be met for molecular
clusters of sufficiently high symmetry, as for the Mn$_6$
molecular cluster compound studied here. The uniaxial anisotropy
of the cluster spin $S = 12$ is as small as $D/k_{B} = 0.013$~K,
giving a total zero-field splitting of the $S = 12$ multiplet of
1.9~K. As a result, the electron-spin lattice relaxation time
remains fast ($\sim 10^{-4}$~s) down to $T_{\rm c}$ and no
blocking occurs. Magnetic specific heat and susceptibility
experiments show a transition to ferromagnetic dipolar order at
$T_{\rm c} = 0.16$~K. Classical Monte Carlo calculations,
performed for Ising $S = 12$ dipoles on a lattice do predict
ferromagnetic ordering and account for the value of $T_{\rm c}$ as
well as the shape of the observed specific heat anomaly. By
applying magnetic fields up to 6~T the hyperfine contributions
$C_{\rm hf}$ to the specific heat arising from the $^{55}$Mn
nuclei could be detected. From the time-dependence of the measured
$C_{\rm hf}$ the nuclear-spin lattice relaxation time
$T_{1\text{n}}$ could be determined for the same field range in
the temperature region $0.2 < T < 0.6$~K. The nuclear magnetic
relaxation was further studied by high field $^{55}$Mn pulse NMR
measurements of both the nuclear $T_{1\text{n}}$ and
$T_{2\text{n}}$ at $T = 0.9$~K (up to 7~T). The data are in good
mutual agreement and can be well described by the theory for
magnetic relaxation in highly polarized paramagnetic crystals and
for dynamic nuclear polarization, which we extensively review. The
experiments provide an interesting comparison with the recently
investigated nuclear spin dynamics in the anisotropic single
molecule magnet Mn$_{12}$-ac.

\end{abstract}

\maketitle

\section{Introduction}

Experimental and theoretical interest in magnetic molecular
clusters carrying a net high spin has rapidly evolved in recent
years (for reviews see, e.g., Refs.
\onlinecite{guntherB,chudnovskyB,gatteschi02B,tupitsyn02B,luis02B}).
Since the cores of these molecules can be viewed as nano sized
pieces of magnetic insulators, they offer attractive possibilities
to study magnetic objects of dimensions in between atom and bulk.
Of great importance  is the fact that these (macro)molecules form
stoichiometric chemical compounds, which may crystallize as
molecular crystals, implying that for a given compound identical
magnetic molecules are arranged on the sites of a regular
three-dimensional lattice. More often than not, there is only a
single molecular site per unit cell, so that the symmetry axes of
all molecules in the crystal are perfectly aligned. Provided that
the intermolecular magnetic interactions are sufficiently weak,
macroscopic solid state techniques can then be exploited to study
the properties of individual cluster spins, taking into account
their couplings to the ``environment'' (phonons, nuclear moments)
as perturbations. Accordingly, such experiments have already
provided highly interesting information about the quantum
tunneling properties of the cluster
spins.\cite{sessoli93N,gatteschi94S,friedman96PRL,Hernandez96,thomas96N,sangregorio97PRL,aubin98JACS,garg93EPL,wernsdorfer99S,mettes01PRB,luis00PRL}

On the other hand, the long-range magnetic ordering (LRMO)
phenomena expected to occur at sufficiently low temperatures as a
consequence of the intercluster magnetic dipolar coupling present
an interesting object of study in
itself.\cite{fernandez00PRB,martinez01EPL} In many of these
compounds, the clusters are only or mainly coupled by van der
Waals forces in the molecular crystal. Short-ranged superexchange
interactions may then be neglected, leaving only the dipolar
coupling between cluster spins as a source for producing LRMO.
Since in the literature of magnetic phase transitions few examples
of LRMO produced purely by dipolar forces are  yet
available,\cite{white93PRL,roser90PRL,noteorder} the chance to
exploit molecular magnets to this end is quite attractive and
could represent an important contribution to this field.

However, for most high-spin molecules studied so far, such as
Mn$_{12}$,\cite{lis80AC} Fe$_8$ \cite{wieghardt84AC}, and the
Mn$_4$ family of
compounds,\cite{aubin96JACS,aliaga01POLY,mettes01POLY} the cluster
spins have strong Ising-type anisotropy, associated with the
zero-field splitting (ZFS) of the magnetic energy levels by the
action of the crystal field. As a consequence, the cluster spins
become frozen below a blocking temperature, $T_{\rm B}$, of
typically a few K, with the spin direction randomly distributed
between the two possible orientations along the easy axis.
Evidently, this superparamagnetic blocking process is in
competition with intercluster magnetic interactions that tend to
establish LRMO at (usually much) lower temperatures. Although it
has been shown theoretically \cite{fernandez02PRB} that the
occurrence of quantum tunneling between opposite spin directions
at temperatures below $T_{\rm B}$  may in principle produce
sufficient fluctuations to overcome blocking, in most of the
investigated anisotropic molecules the times involved for the
actual observation of the ensuing LRMO are still much too long.
The recently discovered example of LRMO found at $T_{\rm c} =
0.21$ K in the molecular magnet Mn$_4$Me is an exception rather
than the rule.\cite{evangelisti04PRL}

\begin{figure}[t]
\includegraphics[width=7.5cm]{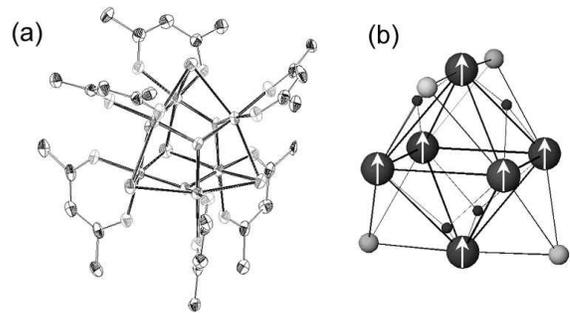}
\caption{\label{structureMn6} (a) Structure of
Mn$_{6}$O$_{4}$Br$_{4}$(Et$_{2}$dbm)$_{6}$; (b) sketch of the
symmetric octahedral core, containing six ferromagnetically
coupled Mn$^{3+}$ ions, yielding a total spin $S = 12$ for this
molecular superparamagnetic particle. The gray and black spheres
indicate the Br$^{-}$ and O$^{2-}$ ions respectively. The white
arrows illustrate the ferromagnetic alignment of the Mn$^{3+}$
$s=2$ spins inside the cluster.}
\end{figure}

The obvious route to find dipolar-induced LRMO in molecular
magnetic cluster compounds is, therefore, to search for high-spin
molecules with as low anisotropy as possible and with negligible
superexchange interactions. In a preliminary report
\cite{morello03PRL} on the compound
Mn$_6$O$_4$Br$_4$(Et$_2$dbm)$_6$  (hereafter called Mn$_6$) we
could show that it provides an excellent example. The Mn$_6$
molecule has a highly symmetric cluster core, comprising an
octahedron of Mn$^{3+}$ ions the faces of which are capped by
O$^{2-}$ or Br$^-$ ions. The structure of the molecule
\cite{aromi99JACS} and a sketch of its octahedral core are shown
in Fig. \ref{structureMn6}. From previous magnetic studies
\cite{aromi99JACS} above $2$ K it was found that the superexchange
paths formed between the Mn$^{3+}$ ions (each having atomic spin
$s = 2$) through the intervening O$^{2-}$ and Br$^-$ ligands
result in a relatively strong ferromagnetic interaction, of value
$J_{\mathrm{f}} /k_{\rm B} \sim +13$ K on basis of the pair
Hamiltonian $\mathcal{H} = - 2 J \mathbf{S}_i \cdot \mathbf{S}_j$.
As a consequence, the ground state is a $S = 12$ multiplet and the
energy of the nearest excited state is approximately $150$ K
higher. The unit cell is monoclinic, with space group $Pc$, and
contains four molecules that have such a high (nearly $T_d$)
symmetry that the net anisotropy for the cluster spin is quite
small. No superexchange paths connecting neighboring clusters can
be discerned indeed in the crystal structure, so that we can
safely assume that the crystal binding arises solely from Van der
Waals bonds. We note that, although intercluster magnetic ordering
has also been reported for the molecular magnets Fe$_{19}$
\cite{affronte02PRB} and Mn$_4$Br,\cite{yamaguchi02JPSJ} in those
cases superexchange between clusters apparently plays an important
role, as evidenced, e.g., by the much higher $T_{\rm c}$ values
found ($1.2 - 1.3$ K). However, for a Cr$_4$S
cluster,\cite{bino88S} for which the intracluster exchange between
the Cr$^{3+}$ ions happens to be likewise ferromagnetic (net spin
$S = 6$), the low value of $T_{\rm c} = 0.17$ K that was observed
could be compatible with dipolar-induced magnetic ordering, in
this case of antiferromagnetic type. More data would be needed,
however, to substantiate this.

In a preliminary report,\cite{morello03PRL} we could already show
that the magnetic anisotropy in Mn$_6$ is sufficiently low to
enable measurements of its magnetic susceptibility and specific
heat under thermal equilibrium conditions down to the lowest
temperatures reached ($T \simeq 15$ mK) by our experimental
setups. The data did evidence a transition to LRMO, as hoped for,
at a temperature of $T_{\rm c} = 0.16$ K, corresponding apparently
to a ferromagnetic arrangement of the cluster moments. Comparison
with Monte Carlo simulations strongly supported the expected
dipolar origin of the intercluster coupling. When applying
magnetic fields of up to $6$ T, the study of the time-dependent
magnetic specific heat revealed a transition from fast relaxation
to non equilibrium conditions within the experimental time window
of  $1 - 100$ s.

In the present work we extend these experimental and theoretical
studies and discuss in much more detail the results and
conclusions. In addition, we have performed $^{55}$Mn NMR studies
in varying field, enabling to draw more definite conclusions about
the magnetic relaxation of both electronic and nuclear spin
systems in this material. The nuclear spin-lattice relaxation time
is governed by fluctuations of the cluster electronic spins, and
is indeed quite fast in zero field. By applying a magnetic field
these fluctuations become progressively suppressed as a
consequence of the Zeeman splitting of the electronic energy
levels, thereby bringing the nuclear spin system out of thermal
equilibrium. This provides an interesting comparison with recent
zero-field $^{55}$Mn NMR studies
\cite{furukawa01PRB,goto03PRB,morello04PRL,morelloT} of the
anisotropic molecular magnet Mn$_{12}$-ac, for which the
suppression of the magnetic relaxation in the thermally activated
regime can be fully ascribed to the strong splitting of the
cluster spin levels by the crystal field. For the present
compound, crystal-field splittings play a very minor role in the
relaxation process, except for providing the necessary channel for
energy transfer between spins and lattice phonons. We present a
full analysis of the longitudinal and transverse nuclear
relaxation in terms of previously developed theories for
relaxation by paramagnetic impurities and for dynamic
polarization, taking into account electron spin fluctuations by
both spin-lattice relaxation and spin-spin interactions. Our NMR
data in high fields prove to be in excellent qualitative as well
as quantitative agreement with such theoretical predictions. The
values for the longitudinal nuclear spin-lattice relaxation rate
and for the effective hyperfine interaction constant deduced from
our high-field time-dependent specific heat data are likewise in
good accord with the NMR results.

The outline of this paper is as follows. After giving a few
experimental details in the next section, the measured
susceptibility and zero-field specific heat data are presented and
discussed in section III A and B, followed by a Monte Carlo
simulation study of the zero-field specific heat in section III C.
In section III D the field-dependent specific heat measurements
are discussed, followed by the nuclear resonance and nuclear
relaxation data in section III E. Section IV contains an analysis
of these data in terms of existing theoretical models for
relaxation in highly polarized magnetic systems. Concluding
remarks are given in section V, while in Appendix A we describe
the calculation of the demagnetizing factor for powder samples.
Systeme International. units will be used throughout the paper.

\section{Experimental details}  \label{paramMn6}

Polycrystalline samples of Mn$_6$ were prepared as reported in
Ref. \onlinecite{aromi99JACS}. Low-temperature specific heat
measurements were performed in a home made calorimeter that makes
use of the thermal relaxation
method.\cite{mettes01PRB,luis02B,mettesT} For the measurements, a
few milligrams of sample were mixed with Apiezon N grease and
placed on the sapphire plate of the calorimeter. Details of this
measurement technique are given in Ref. \onlinecite{mettesT}. An
important advantage of this method is that the characteristic time
$\tau_{\rm e}$ of the experiment (typically, $\tau_{\rm e} \simeq
0.1 - 1000$ seconds at low $T$) can be varied (within limits) by
changing the dimensions (and therefore the thermal resistance) of
the Au wire that acts as a thermal link between the calorimeter
and the mixing chamber of the dilution refrigerator. Magnetic
fields up to 16 T can be applied with a superconducting magnet and
the lowest temperatures reached are of the order of $50$ mK.

The ac-susceptibility measurements were performed between $15$ mK
and $4$ K in a home made susceptometer, placed inside the plastic
mixing chamber of a dilution refrigerator and thermalized by the
$^3$He flow.\cite{morelloT} The susceptometer, based on the mutual
inductance technique, consists of a primary coil with $250$ turns
of $\varnothing 100$ $\mu$m NbTi wire, and two oppositely wound
secondary coils, each with $660$ turns of $\varnothing 40$ $\mu$m
Cu wire. By placing the sample inside one of the two secondary
coils and feeding the primary with an ac current (typically $\sim
50$ $\mu$A), the induced voltage across the secondary is
proportional to the susceptibility of the sample. By
phase-sensitive detection we can also discriminate the real and
imaginary parts of the susceptibility. The excitation frequency
$\nu$ was varied between $230$ and $7700$ Hz. Additional
measurements above $1.8$ K were performed using the ac option of a
commercial Superconducting Quantum Interference Device (SQUID)
magnetometer.

As for the NMR experiments, we introduced the sample and the
four-turns NMR coil inside the plastic $^3$He pot of a pumped
$^3$He cryostat, where variable magnetic fields up to $8$ T could
be applied by a superconducting magnet. The resonance of the
$^{55}$Mn nuclei of Mn$_6$ was observed by means of the spin-echo
technique, with a typical duration of the $\pi/2$ pulse of
$t_{\pi/2} \sim 2 \mu$s.

Estimates of the magnitude of the magnetic anisotropy have been
obtained as follows. Magnetic data measured above $1.8$ K (Ref.
\onlinecite{aromi99JACS}) can be fitted by expressions valid for
fully isotropic $S=12$ spins, i.e. they do not evidence any
detectable ZFS for the total spin. The data are excellently fitted
by the Brillouin curve calculated for an isotropic paramagnet with
$S=12$ and $g=1.98$ [as derived from electron paramagnetic
resonance (EPR)]. If we write a single-spin Hamiltonian for the
molecule as:
\begin{equation}
\mathcal{H} = -DS_z^2 - g \mu_{\rm B} \mathbf{B}_{\rm a} \cdot
\mathbf{S}, \label{hamiltonian}
\end{equation}
these experiments provide an upper limit of $|D|/k_{\rm B}
\lesssim 0.01$ K.\cite{aromi99JACS} In order to obtain an
independent estimation of $D$, high-frequency EPR experiments were
carried out at the NHMFL in Tallahassee by J. Krzystek, using
several frequencies in the range $95$ - $380$ GHz. Simulations of
the spectra performed using Eq. (\ref{hamiltonian}) agree well
with the experimental results for $0.01$ K $<|D|/k_{\rm B}<0.05$
K. A value of $|D|/k_{\rm B} \simeq 0.01$ K seems therefore
appropriate to describe the ZFS in Mn$_6$. When discussing the
ac-susceptibility and specific heat data in sections \ref{suscept}
and \ref{zfspecheat} we shall adopt the value $D/k_{\rm B} =
0.013$ K, which yields the best agreement between theory and
experiment for both techniques.

The isotropic character of the molecular spin might seem
paradoxical at first, considering that the individual Mn$^{3+}$
ions, being Jahn-Teller ions, experience strong anisotropy:
typical values of $|D|/k_{\rm B}$ for the ion are a few tenths of
Kelvin. However, the net $D$ value entering in the spin
Hamiltonian for the cluster can be seen in first approximation to
result from the vectorial addition of the local anisotropy tensors
of the individual ions, which then can give rise to a low net
anisotropy for highly symmetric molecules such as Mn$_{6}$ (cf.
Fig. \ref{structureMn6}), even for large ZFS of the constituting
atoms.\cite{benciniB} In fact, the possibility to tune the net
anisotropy of the cluster spin by means of molecular synthesis is
one of the attractive properties of these nanosized molecular
superparamagnets.

\section{experimental results and analyses}

\subsection{Magnetic susceptibility}\label{suscept}

\begin{figure}[t]
\includegraphics[width=7.5cm]{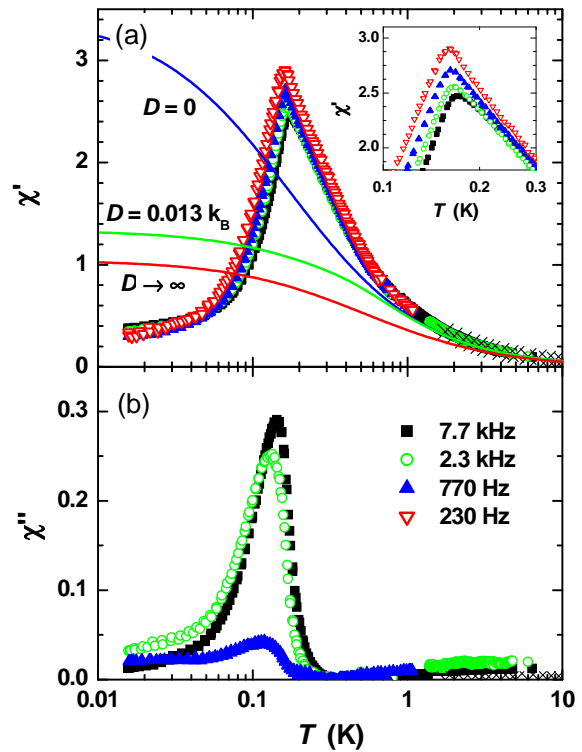}
\caption{\label{X12vsT} (Color online) Real (a) and imaginary (b)
components of the ac-susceptibility at the indicated frequencies.
Inset: magnification of $\chi^{\prime}(T)$ to evidence the
frequency dependence of the peak. The solid lines in panel (a)
give the calculated behavior of the paramagnetic susceptibility
(no interactions) for three values of the crystal field anisotropy
constant $D$. In these calculations, the effect of the
demagnetizing fields of the grains and the sample holder has been
introduced as described in the Appendix.}
\end{figure}

Strong evidence for the long-range ordering of the magnetic
moments is provided by the magnetic susceptibility data, shown in
Fig. \ref{X12vsT}. The real part $\chi^{\prime}$ of the complex
ac-susceptibility is plotted in Fig. \ref{X12vsT}(a), and is seen
to show a sharp maximum at $T_{\rm c} = 0.161(2)$ K. We first
demonstrate that the value of $\chi^{\prime}$ at $T_{\rm c}$ is of
the order of the estimated limit $1/N_{\mathrm{eff}}$ for a
ferromagnetic powdered sample, where $N_{\mathrm{eff}}$ is an
effective demagnetizing factor appropriate for the (cylindrically
shaped) container filled with the grains. In the Appendix we argue
that $N_{\rm eff}$ can be approximately given in terms of the
demagnetizing factors $N_{\rm grain}$ and $N_{\rm cont}$ of,
respectively, the individual grains and the container as [Eq.
(\ref{Neffiso})]:

\begin{equation}
N_{\mathrm{eff}}  = N_{\mathrm{grain}}  + f ( N_{\mathrm{cont}} -
1/3 ),
\end{equation}
where $f$ denotes the volume filling fraction of the container.

Assuming the shape of the grains to be approximately spherical, we
put $N_{\rm grain} = 1/3$, while from the shape of the container,
we estimate $N_{\mathrm{cont}} \simeq 0.2$. The density of the
material is estimated to be $\rho_{\mathrm{grain}} \simeq 1.45$
g/cm$^3$ from the value for the similar compound
Mn$_{6}$O$_{4}$Br$_{4}$(Me$_{2}$dbm)$_{6}$, and the filling
fraction is estimated as $f \sim 1/3$. All this then leads to
$N_{\mathrm{eff}} \simeq 0.29(5)$ and therefore
$\chi^{\prime}(T_{\rm c}) \simeq 1/N_{\rm eff} = 3.5(6)$, the
large error arising obviously from all the uncertainties in the
above line of argument and in the estimates of the parameters
involved. Next we should realize that this value would be valid
for the $\chi^{\prime}$ measured along the easy axis, whereas even
a relatively small anisotropy will lower appreciably the
$\chi^{\prime}$ along the other directions.\cite{steijger84PhyB}
Therefore, the powder $\chi^{\prime}$ could easily be lower by a
factor of 2 [see Eqs. (\ref{suscN1}) and (\ref{suscN2}) of the
Appendix].

Given all the uncertainties, the above derived value can obviously
serve as an order of magnitude estimate only, but we note that the
experimental value of $\chi^{\prime}(T_{\rm c}) \simeq 3$ is
indeed rather close. As a second argument for the ferromagnetic
nature of the transition we include therefore in Fig.
\ref{X12vsT}(a) the  powder susceptibility expected for the
paramagnetic (non interacting)  case, as calculated from the spin
hamiltonian (Eq. \ref{hamiltonian}) and applying in addition the
corrections for demagnetizing effects as described in the
Appendix. Besides the curve for $D/k_{\rm B} = 0.013$ K
appropriate for Mn$_6$, we also include for comparison the fully
isotropic ($D = 0$) and infinite anisotropy limits. From this plot
it is evident that, when approaching $T_{\rm c}$, the
susceptibility of Mn$_6$ increases appreciably above the
paramagnetic limit, confirming the ferromagnetic nature of the
correlations.

Below $T_{\rm c}$, the powder $\chi^{\prime}$ decreases rapidly,
as expected for an anisotropic ferromagnet in which the
domain-wall motions become progressively pinned. The associated
domain-wall losses should then lead to a frequency dependent
maximum around $T_{\rm c}$ in the imaginary part, $\chi^{\prime
\prime}$, as the experimental data of Fig. \ref{X12vsT}(b) indeed
show. In fact, although the Mn$_{6}$ spins can be considered as
nearly isotropic at high temperatures, the anisotropy energy is
still large compared with the dipolar interaction energy $\mu_0
\mu^2 / 4\pi k_{\rm B}r^3 \simeq 0.1$ K between nearest neighbor
molecules. Thus the ordering should be that of an Ising dipolar
ferromagnet.

As shown in detail in the inset of Fig. \ref{X12vsT}(a), the
temperature $T_{\mathrm{peak}}$ at which the maximum value of
$\chi^{\prime}$ is found depends only weakly on $\nu$, which we
attribute to the anisotropy. The total activation energy for the
reversal of each Mn$_{6}$ molecular spin amounts to $DS^{2} \simeq
1.9$ K, i.e., about $35$ times smaller than for Mn$_{12}$.
Although this is quite small, one could still expect the
superparamagnetic blocking of the Mn$_{6}$ spins to occur when $T
\simeq T_{\rm B}($Mn$_{12})/35$, that is below $\simeq 0.15$ K.
Since this value is very close to the actual $T_{\rm c}$, one may
expect that for $T \rightarrow T_{\rm c}$ the approach to
equilibrium begins to be hindered by the anisotropy of the
individual molecular spins. We stress, however, that the frequency
dependence of $\chi^{\prime}$ observed here is quite different
from that of the well-known anisotropic superparamagnetic
clusters. A way to quantify the frequency dependence of the peak
in $\chi^{\prime}$ is by means of the parameter $\Delta
T_{\mathrm{peak}}/[T_{\mathrm{peak}} \Delta (\log_{10} \nu)]$,
which gives the variation of $T_{\mathrm{peak}}$ per decade of
frequency. We find here $\Delta
T_{\mathrm{peak}}/[T_{\mathrm{peak}} \Delta (\log_{10} \nu)]
\simeq 0.03 - 0.05$, to be compared with the typical values of
$\sim 0.20$ for superparamagnetic blocking. In fact it is closer
to the value $\sim 0.06$ found for certain types of spin
glasses,\cite{mydoshB} but the peak observed here is much higher
and sharper. Also, since the cluster spins are situated on a
regular crystal lattice, a comparison with random magnetic systems
would not be appropriate. The frequency dependence we observe is
indeed very much weaker than is found in the
LiHo$_x$Y$_{1-x}$F$_4$ system with $x=0.045$ (Ref.
\onlinecite{reich90PRB}), for which $x$ value that material is in
the ``antiglass'' regime.\cite{reich87PRL,ghosh02S} In Mn$_6$ we
found that $T_{\mathrm{peak}}(\nu) \simeq T_{\mathrm{peak}}(0) + K
\nu^{\alpha}$, with $\alpha \simeq 0.4$ and a zero-frequency limit
of the peak in $\chi'$ $T_{\mathrm{peak}}(0) \simeq 158$ mK. At
essentially the same temperature, we find a fairly sharp peak in
the zero-field magnetic specific heat (see next section), instead
of a broad anomaly as observed in LiHo$_{0.045}$Y$_{0.955}$F$_4$.
Also this finding appears to exclude an interpretation in terms of
a freezing transition in Mn$_6$.

We finally turn to the susceptibility as measured above $T_{\rm
c}$ in the paramagnetic region, which was plotted as
$1/\chi^{\prime}$ \textit{vs} $T$ in the inset of Fig. 2 in Ref.
\onlinecite{morello03PRL}. We first note that no evidence for
relaxation effects were found in this range. Up to a frequency of
$7700$ Hz no appreciable $\chi^{\prime \prime}$ was detected and
the measured $\chi^{\prime}$ smoothly joins the data measured
above $2$ K with the SQUID susceptometer. We may therefore
conclude that in the whole temperature range down to $T_{\rm c}$
the spin-lattice relaxation time is quite short, of the order
$10^{-4}$ s or less.

The high-temperature susceptibility data $\chi^{\prime}_{\rm i}$
corrected for the demagnetizing field ($\chi^{\prime}_{\rm i} =
\chi^{\prime}/[1 - N_{\mathrm{eff}}\chi^{\prime}]$) follow the
Curie-Weiss law $\chi^{\prime}_{\rm i} = C/(T - \theta)$ quite
well down to approximately $0.3$ K, with $C = 0.62(2)$ K and
$\theta= 0.14(3)$ K. The constant $C$ equals, within the
experimental errors, the theoretical value for isotropic spins
$N_{\rm A}g^{2}\mu_0\mu_{\rm B}^{2}S(S+1)/3k_{\rm B}V_{m} = 0.595$
K, where $S = 12$, $g = 2$, and $V_{m} = 1647$ cm$^3$/mole is the
molar volume. The positive $\theta$ confirms the ferromagnetic
nature of the ordered phase. The fact that mean field theory is so
well obeyed down to very close to $T_{\rm c}$ is as expected for a
dipolar
ferromagnet.\cite{roser92PRB,cooke75JPC,beauvillain78PRB,als76PRL}
We remark that the behavior of the powder susceptibility we
observe for Mn$_6$ in Fig. \ref{X12vsT} closely resembles previous
powder data for Cs$_2$NaGdCl$_6$ (cf. Fig.3 of
\onlinecite{roser92PRB}), and LiHo$_{x}$Y$_{1-x}$F$_4$ ($x=0.46$)
(cf. Fig.1 in \onlinecite{reich86PRB} and Fig. 4 in
\onlinecite{reich90PRB}). Both materials are considered to be
examples of anisotropic dipolar ferromagnets. Although weak
ferromagnetism (i.e., canted antiferromagnetism) would also lead
to $\chi(T=T_{\rm c}) = 1/N$, this would be accompanied by a
negative Curie-Weiss $\theta$, whereas we observe a positive
value. Moreover, for canted antiferromagnetism an antisymmetric
interaction term of the form $\mathbf{d} \cdot(\mathbf{S}_i \times
\mathbf{S}_j)$ is needed (see e.g. Ref. \onlinecite{moriya63B})
and this is not expected for a system of equivalent magnetic
moments interacting by dipolar interactions, as is the case here.
For metamagnetic systems, having ferromagnetic nearest neighbor
interactions and weaker further neighbor couplings, also a
positive value for $\theta$ can be found but the magnetic ordering
below $T_{\rm c}$ due to the further neighbor interaction is
basically antiferromagnetic. Consequently, although the value of
$\chi(T=T_{\rm c})$ can be much higher than for nearest neighbor
antiferromagnets, it will fall still far below the ferromagnetic
limit of $1/N$. We therefore conclude that the evidence for
ferromagnetic dipolar order in Mn$_6$ presented here is quite
strong. Additional studies, for instance of the spontaneous
magnetization below $T_{\rm c}$, would of course be welcome to
provide further proof.

\subsection{Zero-field specific heat}\label{zfspecheat}

\begin{figure}[t]
\includegraphics[width=7.5cm]{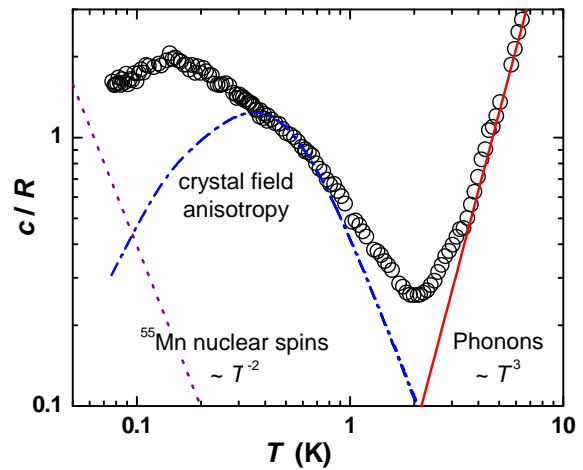}
\caption{\label{CmvsTB0} (Color online) Zero-field specific heat
of Mn$_6$ plotted as a function of temperature. Solid line: phonon
contribution ($C/R=0.010$ $T^3$). Dashed-dotted line: Schottky
contribution due to crystal field splitting of the $S=12$
multiplet as calculated with $D/k_{\rm B} = 0.013$ K. Dashed
curve: expected nuclear contribution from the $^{55}$Mn nuclear
spins.}
\end{figure}

Specific heat data $c$ taken in zero applied field are shown in
Fig. \ref{CmvsTB0} as a function of temperature on a double
logarithmic scale. Above $2$ K, the specific heat is dominated by
the contribution from the lattice phonons that can be reasonably
fitted by the well-known low-$T$ Debye approximation:
$c_{\mathrm{latt}} \propto (T/\Theta_D)^3$, with a Debye
temperature $\Theta_D \simeq 29$ K. Such low values are commonly
observed in the molecular cluster
compounds,\cite{mettes01PRB,evangelisti04PRL,luis02B} reflecting
the weak bonding between the cluster molecules in such molecular
solids.

\begin{figure}[t]
\includegraphics[width=7.5cm]{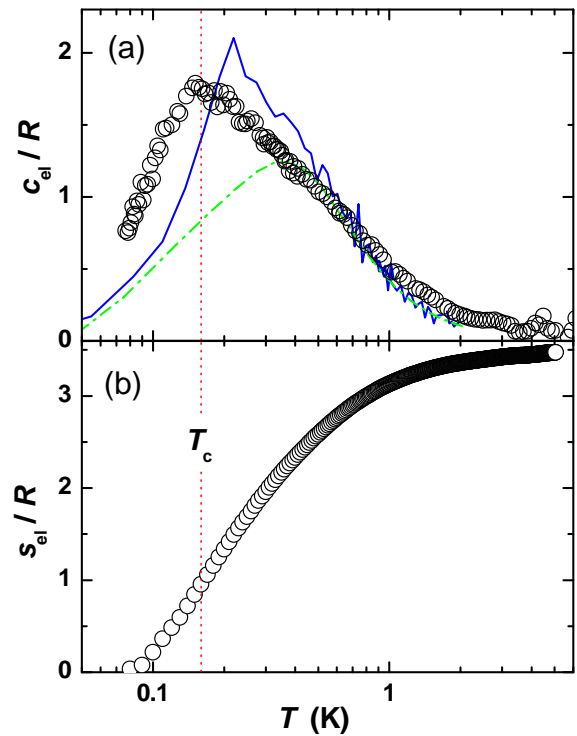}
\caption{\label{CevsT} (Color online) (a) $T$ dependence of the
electronic specific heat at zero applied field (circles). The
dashed-dotted line is the Schottky anomaly calculated with
$D/k_{\rm B}=0.013$ K. The solid line is the Monte Carlo (MC)
calculation for an orthorhombic lattice of $1024$ Ising spins with
periodic boundary conditions. For each point, $2 \times 10^{4}$ MC
steps per spin were performed. (b) $T$ dependence of the entropy
obtained by integration of the electronic specific heat curve.}
\end{figure}

In the lowest temperature range below $0.1$ K, the specific heat
is seen to remain rather high valued, which can be mainly ascribed
to the expected contribution from the $^{55}$Mn nuclear spins, the
energy levels of which are split by the hyperfine interaction with
the Mn$^{3+}$ electronic spins $s=2$ (see section III D below).
This contribution will become more clearly and directly visible in
the field-dependent studies, where it will be shown to be
describable by a term $c_{\mathrm{nucl}}T^2/R \simeq 5.2 \times
10^{-3}$ K$^2$, which is plotted as the dashed line in Fig.
\ref{CmvsTB0}. Anticipating the discussion below, we subtract this
hyperfine contribution, as well as the phonon $\propto T^3$ term
from the raw data in Fig. \ref{CmvsTB0}, in order to obtain the
magnetic specific heat $c_{\mathrm{el}}$ associated with the
cluster electron spin $S=12$ only, as plotted in Fig. \ref{CevsT}.
The resulting curve shows two characteristic features. At a
temperature $T_{\rm c} \simeq 0.15(2)$ K a peak is observed, that
can be associated with the transition to long-range magnetic
order, in good agreement with the value for $T_{\rm c}$ deduced
from the peak in the ac-susceptibility extrapolated to zero
frequency. Above $T_{\rm c}$, one observes a widely extended
``high-temperature tail'', that reflects the weak zero-field
splittings (ZFS) of the $S=12$ multiplets by the crystal field
interactions. We recall that even for a $D$ value as small as
$0.013$ K the total splitting of an $S=12$ multiplet will still be
an appreciable $DS^2 = 1.9$ K. In the absence of magnetic
intercluster interactions the ZFS of such a multiplet would lead
to a multilevel Schottky curve, shown as the dashed-dotted curve
in both Fig. \ref{CmvsTB0} and \ref{CevsT}, where the fit to the
experiment leads to the estimate $|D|/k_{\rm B} \simeq 0.013$ K,
in reasonably good agreement with the values quoted above. In
contrast with the highly anisotropic molecular clusters Mn$_{12}$,
Mn$_4$, and Fe$_8$, where the multilevel ZFS Schottky is found
above $1$ K and is the most pronounced feature of the
experimentally observed magnetic specific
heat,\cite{mettes01PRB,evangelisti04PRL,luis02B} its presence in
Mn$_6$ is masked at the low-$T$ side by the ordering anomaly
produced by the effects of the intercluster magnetic interactions.
Numerical integration of $c_{\mathrm{el}}/T$ between $0.08$ K and
$4$ K gives a total entropy change $\Delta s_{\rm el}/R = 3.5(2)$
per mole, close to the expected total entropy for a fully split
$S=12$ multiplet, namely $\Delta s_{\rm el}/R = \ln(2S+1) = 3.22$.
This confirms the consistency of the above subtraction procedure
used to obtain $c_{\rm el}$. The variation with temperature of the
entropy, $s_{\rm el}/R$, is also shown in Fig. \ref{CevsT}. We
note in particular that at $T_{\rm c}$ itself, the entropy only
amounts to about $1 R$ per mole, indicating that only the lowest
energy levels of the cluster spins are involved in the actual
magnetic ordering process, the majority of the higher-lying levels
being already depopulated (the lowest lying doublet ground state
on its own would already give a contribution to the entropy of $R
\ln 2 = 0.69 R$ per mole). Indeed, the distance to the nearest
lying excited state would be equal to $(2S-1)D \simeq 0.3$ K in
the non interacting limit.

\subsection{Monte Carlo simulations}

To simulate the zero-field specific heat data, Monte Carlo (MC)
calculations were performed for an $S=12$ Ising model of magnetic
dipoles on a body-centered orthorhombic lattice containing $Z=2$
molecules, with axes $a_x = 15.7$ \AA, $a_y=23.33$ \AA, and
$a_z=16.7$ \AA. This choice approximates the crystal structure of
Mn$_{6}$O$_{4}$Cl$_{4}$(Et$_{2}$dbm)$_{6}$, that is the most
closely related compound for which the structure could be
resolved. To further simplify the Monte Carlo simulations, we
approximated the monoclinic structure by an orthorhombic one. The
density, $\rho = 1.45$ g/cm$^3$, was estimated from the
Mn$_{6}$O$_{4}$Br$_{4}$(Me$_{2}$dbm)$_{6}$ compound. The resulting
molar volume is a few percent larger than what would be obtained
from the lattice parameters used in the Monte Carlo simulation,
but this discrepancy is due entirely to the approximation of
orthorhombic structure. The Hamiltonian includes the dipolar
interaction term as well as the anisotropy term $-DS_z^2$ given in
Eq. (\ref{hamiltonian}). For $T < 0.5$ K the intermolecular
dipolar interactions become important and remove the degeneracy of
the $|\pm m\rangle$ spin doublets. The MC simulations show that
the ground state is indeed ferromagnetically ordered, as observed,
and predict a shape for $c_{\rm el}$ that is in good agreement
with the experiment. In the upper panel of Fig. \ref{CevsT}, we
show $c_{\rm el}$ as calculated assuming all molecular easy ($z$)
axes to point along $a_z$, i.e., one of the two nearly equivalent
short axes of the actual lattice. Similar results were obtained
for other orientations chosen for the anisotropy ($z$) axis. We
note that our simulations give $T_{\rm c} = 0.22$ K, which is
slightly higher than the experimental $T_{\rm c}=0.161(2)$ K. This
difference may be due to the Ising approximation taken for the
intercluster dipolar interaction and to uncertainties in the
values of the lattice parameters. In fact we note in passing that
an almost perfect coincidence of our calculated curve with the
experimental data may be obtained by assuming a smaller value of
the magnetic moment, i.e., taking $g=1.75$ instead of $g=2.00$.

\begin{figure}[t]
\includegraphics[width=7.5cm]{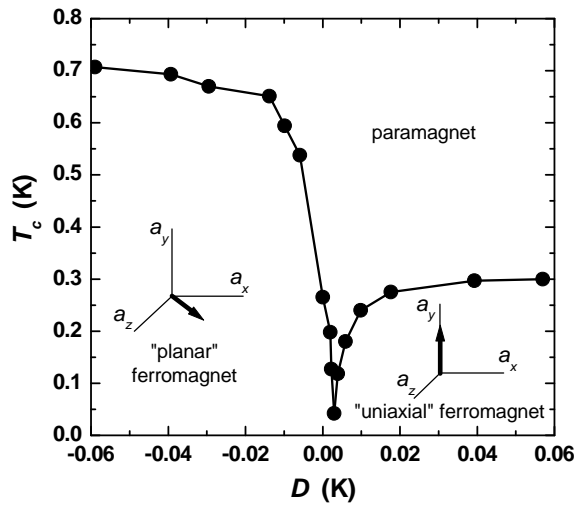}
\caption{\label{TcvsD} Calculated critical temperature $T_{\rm c}$
as a function of the anisotropy parameter $D$, for classical
Heisenberg spins.}
\end{figure}

To pursue this point further, additional MC simulations were
performed for the same crystal lattice, but now with classical
Heisenberg spins replacing the $S=12$ Ising spins. To investigate
the sensitivity of our results to the type of anisotropy, the sign
and magnitude of $D$ were varied. These calculations resulted in
the phase diagram shown in Fig. \ref{TcvsD}, which we include here
since it illustrates the complicated way in which the nature of
the actual ground state and the value of $T_{\rm c}$ may depend on
the combination of the long-range dipolar interaction and the
anisotropy parameter. Although the ground state for this lattice
is always found ferromagnetic, it can be either ``uniaxial'', with
strong preference for the spins to lie along the  $a_y$ axis
[chosen for this particular example to be the anisotropy axis $z$
of Eq. (\ref{hamiltonian})], or ``planar'', in the sense that the
$a_x - a_z$ plane becomes an easy plane, with a weak preference
for a given direction in the plane, as sketched in the figure.
Interestingly, the switching point between these ``uniaxial''and
``planar'' orientations is not at $D=0$. The reason for this is as
follows. Because the crystal lattice under consideration is far
from being cubic, the dipolar interaction energy is rather
anisotropic. The dipolar energy is minimized when the
magnetization $\mathrm{M}$ points in the direction (in the $a_x -
a_z$ plane) shown in the inset on the left-hand side of Fig.
\ref{TcvsD}. Therefore, $\mathrm{M}$ points along this direction
for $D<0$. On the other hand, the energy minimization for $D>0$ is
a competing process. Clearly, dipolar interaction must become
dominant for sufficiently small values of $D$. The numerical
results show that this occurs if $0 < D/k_{\rm B} \lesssim 3$ mK.
The numerical datapoints in Fig. \ref{TcvsD} also show that
$T_{\rm c}$ varies sharply within the $-0.01 < D/k_{\rm B} < 0.01$
K range. System size effects and computer time restrictions do not
allow us to determine whether $T_{\rm c}$ vanishes completely. The
lowest numerical value obtained is as small as $T_{\rm c} \simeq
0.03$ K at $D/k_{\rm B} \simeq 3$ mK. Outside this narrow range of
$D$, $T_{\rm c}$ is already almost equal to the limiting values of
$\simeq 0.7$ K and $\simeq 0.3$ K, reached for infinite negative
and positive $D$, respectively. Such a variation of $T_{\rm c}$
with anisotropy, as well as the form of the calculated and
observed specific heat ordering anomaly, appear to be specific for
dipolar interactions. They differ widely from the corresponding
behavior known for three-dimensional (Heisenberg, Ising, XY)
ferromagnetic lattices with nearest-neighbor interaction
only.\cite{dejongh01AP} For instance, in those models the
variation of $T_{\rm c}$ with anisotropy is restricted to about
20\% of the $T_{\rm c}$ value, which moreover is highest for the
Ising (``uniaxial'') case. It is also interesting to compare these
Monte Carlo calculations with the predictions of simple mean field
theory. The latter gives $T_{\rm c} = 2S^{2} J_{\rm eff}/k_{\rm
B}$ and $T_{\rm c} = 2S(S+1)J_{\rm eff}/3 k_{\rm B}$, respectively
for $D \rightarrow + \infty$ and $D=0$, where $J_{\rm eff}$ is an
effective interaction constant. The ratio between these two
limits, which is $3S/(S+1)$, is about 2.5 times larger than what
is obtained by Monte Carlo calculations.

In concluding this section we may stress that our detailed
calculations have evidenced that for a dipolar magnet the value of
$T_{\rm c}$ may vary strongly with anisotropy and lattice
symmetry. This illustrates the danger in drawing conclusions about
the nature of the magnetic interactions, i.e., whether they are of
dipolar origin or not, just by comparing the value of $k_{\rm B}
T_{\rm c}$ with the dipolar interaction energy of a pair of
nearest neighboring spins. Our detailed calculations
\emph{specific} for this compound show a good quantitative
agrement assuming just dipolar coupling, both as regards the value
of $T_{\rm c}$ and the shape of the specific heat anomaly.
Furthermore, the prediction that the ordering is ferromagnetic, as
observed, is quite robust since it is independent of the details
of the simulations. All this confirms that LRMO in Mn$_6$ is
mainly driven by dipolar interactions.

\subsection{Field-dependent specific heat: Nuclear spin contribution}  \label{sec:CmvsB}

\begin{figure}[t]
\includegraphics[width=7.5cm]{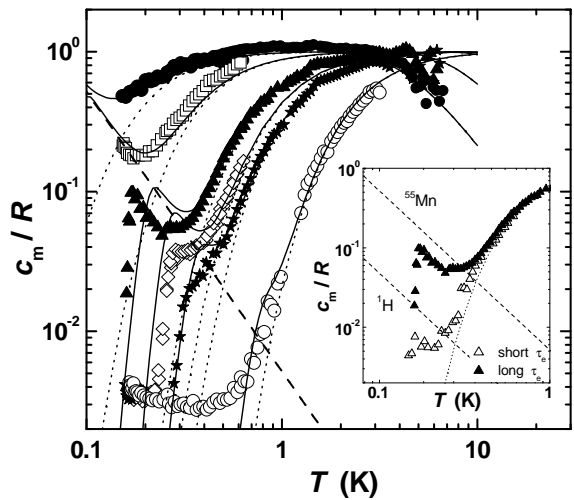}
\caption{\label{CmvsT-B} Temperature dependence of the magnetic
specific heat $c_{\rm m}$ at $B_{\rm a}=0.5$ T ($\bullet$), $1$ T
($\square$), $2$ T ($\blacktriangle$), $2.5$ T ($\diamond$), $3$ T
($\star$), and $6$ T ($\circ$). The dotted lines represent the
calculated electronic contribution, whereas the dashed line is the
expected nuclear specific heat at zero field calculated with Eq.
(\ref{cnucleq}). The solid lines represent the total
time-dependent $c_{\rm m}$ (electronic + nuclear) calculated
accounting for the nuclear $T_{\mathrm{1n}}$ (see text). Inset:
detail of $c_{\rm m}(T)$ at $B_{\rm a}=2$ T, for long ($\sim 100$
s) and short ($\sim 1$ s) experimental times. The dashed lines are
the calculated contributions arising from the $^{55}$Mn and
$^{1}$H nuclear spins.}
\end{figure}

We next discuss the time-dependent magnetic specific heat $c_{\rm
m} = c - c_{\rm latt}$ measured under varying applied magnetic
fields $B_{\rm a}$, as plotted in Fig. \ref{CmvsT-B}. Even for the
lowest $B_{\rm a}$ value, the ordering anomaly is already fully
suppressed, as expected for a ferromagnet. Accordingly, we may
account for these data with the Hamiltonian (\ref{hamiltonian})
neglecting dipolar interactions. The Zeeman term splits the
otherwise degenerate $| \pm m \rangle$ doublets, and already for
$B_{\rm a} \sim 0.5$ T the level splittings become predominantly
determined by $B_{\rm a}$, so that the anisotropy term can then
also be neglected. As seen in Fig. \ref{CmvsT-B}, the calculations
performed with $D=0$ (dotted curves) reproduce quite
satisfactorily the data at higher temperatures.

However, when the maxima of the multilevel Schottky anomalies are
shifted to higher $T$ by increasing $B_{\rm a}$, an additional
contribution is revealed at low $T$. It is most clearly visible in
the curves for $1$ T $< B < 2.5$ T, and varies with temperature as
$c_{\rm m}T^{2}/R \simeq 4 \times 10^{-3}$. We attribute this
contribution to the already-mentioned high-temperature tail of the
equilibrium nuclear specific heat $c^{\rm (eq)}_{\rm nucl}$. As
discussed in section IV, this specific heat should be dominated by
the contributions of the six $^{55}$Mn nuclear spins ($I=5/2$) and
of the $114$ protons present in each molecule. The energy levels
of the former are split, even at zero field, by the strong on-site
hyperfine interactions with the Mn$^{3+}$ electronic spins $s=2$.
For the specific heat analysis this interaction can be
approximated by $\mathcal{H}_{\mathrm{hf}}=A \mathbf{I} \cdot
\mathbf{s}$, where $A$ is an effective isotropic hyperfine
coupling constant. By contrast, we may neglect the hyperfine
splitting of the protons because it can be expected to be small
compared to their nuclear Zeeman splitting for $B_{\rm a} > 1$ T.
This is indeed confirmed by the NMR experiments shown below. The
high-temperature limit of the nuclear specific heat can therefore
be approximated by the sum of two contributions:\cite{abragam70}
\begin{eqnarray}
\frac{c^{\rm (eq)}_{\mathrm{nucl}} T^{2}}{R} \simeq 6 \times
\frac{1}{3}A^{2}s^{2}I(I+1) + 114 \times \left (\frac{\hbar
\gamma_{\rm H} B_{\rm a}}{2 k_{\rm B}} \right)^{2} \label{cnucleq}
\end{eqnarray}

\noindent where $\gamma_{\rm H} = 2.675 \times 10^8$ rad T$^{-1}$
s$^{-1}$ is the protons gyromagnetic ratio. Taking $A/k_{\rm
B}=8.6$ mK as estimated from the NMR spectra measured for the same
sample (see section III E), we obtain the dashed line shown in
Fig. \ref{CmvsT-B} at zero field. This is the same contribution
that was subtracted from the zero-field data shown in Fig.
\ref{CevsT}. The difference between the calculated and
experimental $c^{\rm (eq)}_{\rm nucl}$, which becomes especially
evident for the $B_{\rm a}=2$ T curve, can be due to the shift of
the $^{55}$Mn nuclear energy splittings by the applied field,
which is neglected in Eq. (\ref{cnucleq}).

A remarkable feature of the experimental data that is not
reproduced by these equilibrium calculations is that, at the
lowest $T$, the nuclear specific heat drops abruptly to a baseline
of about $3 \times 10^{-3} R$. This remaining specific heat is
probably a background feature arising from incomplete correction
for the field-dependent addenda contributions. The crossover
temperature $T^{*}$ where the drop of $c_{\rm m}$ occurs depends
on $B_{\rm a}$ but also on the characteristic time constant
$\tau_{\rm e}$ of our (time-dependent) specific heat experiment:
as is shown in the inset of Fig. \ref{CmvsT-B} the deviation from
the (calculated) equilibrium specific heat is found at a lower $T$
when the system is given more time to relax. Interestingly, the
specific heat becomes even smaller than the expected contribution
of the protons. The drop therefore shows that, below $T^{*}$, the
nuclear spins of both the $^{1}$H and $^{55}$Mn atoms cannot
attain thermal equilibrium with the lattice phonons within the
experimental time $\tau_{\rm e}$ because the longitudinal nuclear
spin-lattice relaxation time becomes too short. This relaxation
effect can be described as follows:
\begin{eqnarray}
c_{\rm nucl}(\tau_{\rm e})= c_{\rm nucl}^{(\mathrm{eq})}
[1-\exp(-\tau_{\rm e}/T_{1\text{n}})],  \label{CnuclT1}
\end{eqnarray}
\noindent where, to simplify the discussion, we have used the same
nuclear spin-lattice relaxation (NSLR) time $T_{1\text{n}}$ for
both protons and $^{55}$Mn although they can obviously differ from
each other. According to Eq. (\ref{CnuclT1}) the specific heat
decreases fast when $T_{1\text{n}}$ becomes of the order of
$\tau_{\rm e}$. This crossover to a non equilibrium regime (as
measured by time-dependent specific heat) provides therefore
direct information on the temperature and field dependence of the
nuclear $T_{1\text{n}}$.

\begin{figure}[t]
\includegraphics[width=7.5cm]{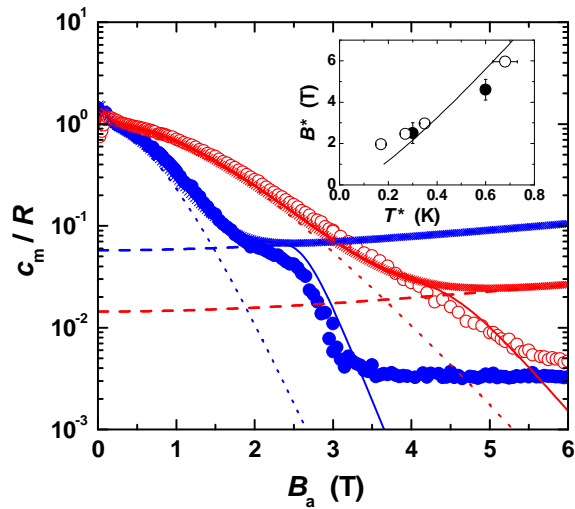}
\caption{\label{CmvsB} (Color online) Field dependence of the
magnetic specific heat at $T=0.3$ K (solid dots) and $T=0.6$ K
(open dots). The dashed and dotted lines represent respectively
the calculated nuclear (see Eq. (\ref{cnucleq})) and electronic
contributions to the equilibrium specific heat. The thick and thin
solid lines give respectively the equilibrium $c_{m}$ and the
time-dependent $c_{\rm m}$ calculated accounting for the
field-dependent nuclear $T_{1\text{n}}$. Inset: $B^{*}(T^{*})$
obtained from $T$ sweeps, as in Fig. \ref{CmvsT-B} (open dots), or
from $B$ sweeps, as in the present figure (solid dots). From the
fit (solid line) to Eq. (\ref{T1Mn6}) we extract $\kappa_{0}
\simeq 40$ s$^{-1}$T$^{-3}$.}
\end{figure}

As is well established,\cite{abragam61} $T_{1\text{n}}(T,B_{\rm
a})$ can be related to the time-dependent part of the transverse
hyperfine field as produced by the fluctuations of the electron
spin. For the case of Mn$_{6}$ both the electron spin fluctuations
due to spin-lattice coupling and to dipolar spin-spin interactions
will have to be considered. A more extensive theoretical treatment
is given below in section IV, in terms of existing models for
nuclear relaxation in magnetic crystals. As will be seen, this
treatment predicts the behavior of the longitudinal nuclear
spin-lattice relaxation time at low temperature and high fields
($>3$~T) to be given by
\begin{equation}
\frac{1}{T_{\rm 1n}} \approx \kappa_{0} B_{\rm
a}^{3}\exp\left(-\frac{g \mu_{\rm B} B_{\rm a}}{k_{\rm B} T}
\right) \label{T1Mn6} \ ,
\end{equation}
where $\kappa_{0}$ is a constant that depends on the electronic
spin-lattice relaxation rate and on the details of the relaxation
mechanism. This shows that an exponential temperature dependence
of the NSLR rate is expected at high fields. The effect of the
field is to polarize the electron spins, which reduces the
fluctuations of the hyperfine field, thus effectively
disconnecting the nuclear spins from the lattice.

As seen from Eq. (\ref{T1Mn6}) the nuclear spins can be taken out
of equilibrium either by decreasing $T$ down to $T^{*}$ at
constant field (as in Fig. \ref{CmvsT-B}), or by increasing
$B_{\rm a}$ up to a given value $B^{*}$ at constant $T$. The
latter effect is indeed also observed experimentally, as shown in
Fig. \ref{CmvsB} for $T=0.3$ K and $0.6$ K. In this figure the
transition to non equilibrium is obvious from the fact that the
data measured in high field fall far below the expected nuclear
contributions (dashed lines). The inset of Fig. \ref{CmvsB} shows
$B^{*}(T^{*})$ obtained either from $T$ sweeps at constant $B$ (as
in Fig. \ref{CmvsT-B}) or from $B$ sweeps at constant $T$ (as in
Fig. \ref{CmvsB}). The two methods prove to be fully consistent
with each other. The fit of $B^{*}(T^{*})$ using Eq. (\ref{T1Mn6})
gives an average value of $\kappa_{0} \simeq 40$~s$^{-1}$T$^{-3}$.
Using this value of $\kappa_0$ we have calculated the
time-dependent $c_{\mathrm{nucl}}$ from Eq. (\ref{CnuclT1}).
Adding this to the calculated electronic specific heat yields the
solid lines in Figs. \ref{CmvsT-B} and \ref{CmvsB}, which can be
seen to be in reasonably good agreement with the experimental data
over the whole range of field and temperature.

\subsection{$^{55}$Mn - NMR and nuclear relaxation}

The nuclear spin-lattice relaxation in Mn$_6$ has been further
investigated by $^{55}$Mn nuclear magnetic resonance. Apart from
the fundamental question as to by which mechanisms the magnetic
relaxation proceeds in this isotropic molecular magnetic crystal,
it is of interest to compare with the typical behavior observed
recently for the highly anisotropic single-molecule magnets like
Mn$_{12}$-ac.\cite{furukawa01PRB,goto03PRB,morello04PRL}

As is well known, it is difficult to observe NMR for nuclei of
paramagnetic ions due to the very large and strongly fluctuating
magnetic fields produced at the nuclei by the electron spin
through the (on-site) hyperfine interactions. As a consequence,
nuclear resonance lines become very broad and spin-lattice
relaxation rates too fast to be measured. To enable the
observation of the NMR signals, one should therefore take recourse
to the low-temperature and high-field regime, in which electron
spin fluctuations can be expected to be sufficiently suppressed.
Accordingly, we performed our experiments at $T=0.9$~K, using a
$^{3}$He cryostat, and fields in the range 3 to 7 T.

\begin{figure}[t]
\includegraphics[width=7.5cm]{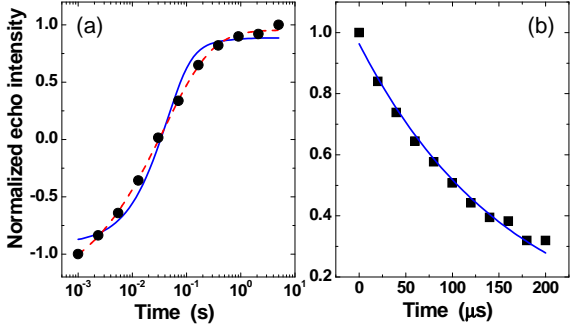}
\caption{\label{T1T2Mn6} (Color online) Inversion recovery (a) and
decay of transverse magnetization (b) for the $^{55}$Mn nuclei at
$T=0.9$ K, $B_{\rm a} = 5$ T and $\nu = 251.5$ MHz. The lines in
(a) are fits to Eq. (\ref{strexp}) with $\alpha \simeq 0.5$
(dashed) and $\alpha =1$ (solid). The solid line in (b) is a fit
to Eq. (\ref{T2}).}
\end{figure}

The $^{55}$Mn NSLR was studied by measuring the recovery of the
nuclear magnetization after an inversion pulse. By integrating the
echo intensity we obtained recovery curves as those shown in Fig.
\ref{T1T2Mn6}(a). For the ease of comparison between different
curves, we renormalize the vertical scale such that
$M(0)/M(\infty)=-1$ and $M(t \gg T_{1\text{n}})/M(\infty)=1$, even
though usually $|M(0)|<|M(\infty)|$: this is just an artifact that
occurs when the NMR line is much broader than the spectrum of the
inversion pulse, and does not mean that the length of the $\pi$
pulse is incorrect. Since the $^{55}$Mn nuclei have spin $I =
5/2$, the recovery of the nuclear relaxation for the central line
in the quadrupolar split manifold is described
by\cite{suter98JPCM}
\begin{widetext}
\begin{eqnarray}
    \frac{M(t)}{M(\infty)} = 1 - \left[ \frac{100}{63} e^{-(15t/T_{1\text{n}})^{\alpha}}
     +\frac{16}{45} e^{-(6t/T_{1\text{n}})^{\alpha}} + \frac{2}{35} e^{-(t/T_{1\text{n}})^{\alpha}} \right],
\label{strexp}
\end{eqnarray}
\end{widetext}
where $1/T_{1\text{n}}$ is the nuclear spin-lattice relaxation
rate, and $\alpha$ is a stretching exponent, which needs to be
introduced to account for the large inhomogeneity of the NMR line,
which causes the inversion recovery to consist of a combination of
recoveries with different rates. We typically found an optimal
value of $\alpha \simeq 0.5$, although the choice of the
stretching exponent does not strongly influence the value of
$1/T_{1\text{n}}$ extracted from the fit.

The transverse spin-spin relaxation (TSSR) rate $1/T_{2\text{n}}$
was studied by measuring the decay of echo intensity upon
increasing the waiting time $\tau$ between the $\pi/2$- and the
$\pi$-pulses. The decay of transverse magnetization
$M_{\perp}(\tau)$ can be fitted to a single exponential:
\begin{eqnarray}
\frac{M_{\perp}(2\tau)}{M_{\perp}(0)}=e^{-2\tau/T_{2\text{n}}},
\label{T2}
\end{eqnarray}
as shown in Fig. \ref{T1T2Mn6}(b).

\begin{figure}[t]
\includegraphics[width=7.5cm]{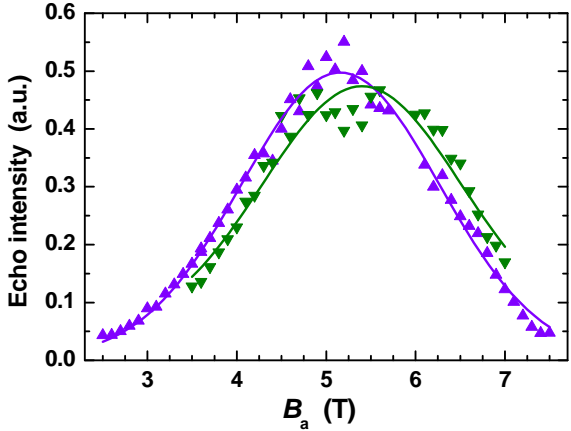}
\caption{\label{spectrumMn6} (Color online) $^{55}$Mn NMR spectra
at $T=0.9$ K and measuring frequencies $\nu=246.5$ MHz (down
triangles) and $\nu=251.5$ MHz (up triangles). The gap in the data
around $B_{\rm a} \simeq 5.8$ T is due to the crossing with the
$^{1}$H line arising from the Zeeman-split proton levels. The
lines are Gaussian fits with total width $2 \sigma_B \simeq 2.2$
T.}
\end{figure}

The field-sweep NMR spectra in Fig. \ref{spectrumMn6} clearly show
that it is impossible to determine whether there are inequivalent
sites in the molecule, as regards the hyperfine coupling [compare
with the case of Mn$_{12}$-ac (Ref. \onlinecite{kubo02PRB})]. This
may be due to the large quadrupolar splitting expected in
Mn$^{3+}$ sites, plus the fact that our sample is an unoriented
powder. As expected from the internal ferromagnetic structure of
the cluster electron spins, the $^{55}$Mn spectrum shifts to
higher fields when lowering the frequency. The spectra can be
fitted by a Gaussian shape with total width $2 \sigma_B \simeq
2.2$ T. If this width were due to quadrupolar splitting $\Delta
\nu_Q$ only, one would deduce $\Delta \nu_Q \sim 7$ MHz: this
estimate can be obtained from a comparison to the Mn$^{(1)}$ line
in Mn$_{12}$-ac, where $\Delta \nu_Q = 0.72$ MHz yields $2
\sigma_{\nu} = 2.4$ MHz,\cite{kubo02PRB} i.e. $2 \sigma_B = 2
\sigma_{\nu} / (\gamma_{\rm Mn} / 2 \pi) \simeq 0.23$ T, where
$\gamma_{\rm Mn} = 6.64 \times 10^7$ rad T$^{-1}$ s$^{-1}$ is the
$^{55}$Mn gyromagnetic ratio. Such an estimate is thus even larger
than the highest $\Delta \nu_Q \simeq 4.3$ MHz found in the
Mn$^{(2)}$ sites of the less symmetric Mn$_{12}$-ac
cluster.\cite{kubo02PRB} We expect therefore that the random
orientation of the crystallites and, eventually, the presence of
inequivalent Mn sites as regards the hyperfine coupling, are also
contributing to the observed broadening. Indeed, when decreasing
the frequency by 5 MHz the maximum of the spectrum shifts only by
$0.24$ T, instead of the $0.47$ T that would be expected when all
the local hyperfine fields are antiparallel to $\mathrm{B}_{\rm
a}$. We conclude that the observed spectrum, as well as the NSLR
and TSSR data, should be considered as obtained from a mixture of
nuclear signals arising from randomly oriented crystallites with
largely overlapping and quadrupolar-split NMR lines from all the
Mn sites in the cluster. Extrapolating to $B_{\rm a}=0$ the field
dependence of the peak of the spectrum, one obtains $\nu(0) \simeq
360$ MHz $\Rightarrow B_{\mathrm{hyp}} \simeq 34$ T, very similar
to the highest value found in the Mn$^{(3)}$ site of
Mn$_{12}$-ac.\cite{kubo02PRB} Finally, in connection with the
discussion of the nuclear specific heat of the previous section,
it is interesting to mention that the $^{1}$H resonance is found
at a value of field given simply by $B_{\rm H} \simeq \omega /
\gamma_{\rm H}$ (the excluded region in the spectra shown in Fig.
\ref{spectrumMn6}). This confirms that the local hyperfine fields
do not appreciably shift the $^{1}$H resonance frequency and
therefore, for sufficiently high fields ($B_{\rm a} > 1$ T), the
hyperfine interaction of protons can be neglected, as we did.

\begin{figure}[t]
\includegraphics[width=7.5cm]{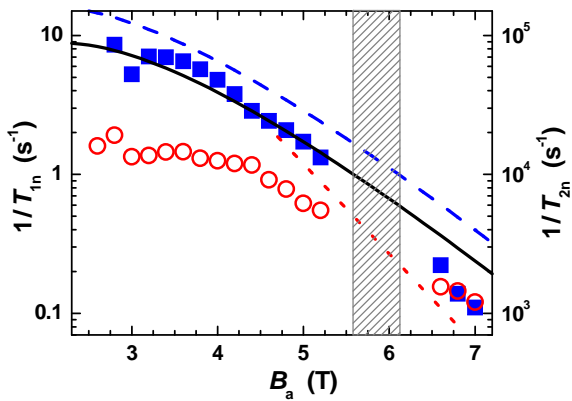}
\caption{\label{W1W2Mn6} (Color online) Field dependencies of the
$^{55}$Mn NSLR (full squares, left scale) and TSSR (open circles,
right scale) at $T=0.9$ K and $\nu=251.5$ MHz. Notice the factor
$10^4$ between left and right scales. The hatched area around 5.8
T indicates the region where the $^1$H line overlaps with the
$^{55}$Mn resonance. Dashed line: calculated $1/T_{1 {\rm
n}}(B_{\rm a})$ according to Eq. (\ref{T1Mn6}) for the value
$\kappa_{0} = 40$ s$^{-1}$T$^{-3}$ as extracted from specific heat
experiments. This represents exactly the same line as in the inset
of Fig. \ref{CmvsB}. Solid line: fit to Eq. (\ref{T1star}) below
with parameters discussed in the text. Dotted line: $1/T_{\rm 2n}$
calculated according to Eq. (\ref{F.13}).}
\end{figure}

Figure \ref{W1W2Mn6} shows the field dependencies of the NSLR rate
$1/T_{1\text{n}}$ and the TSSR rate $1/T_{2\text{n}}$, measured at
constant frequency $\nu = 251.5$ MHz and temperature $T=0.9$ K.
From the discussion above it is clear that these data must be
interpreted with a certain caution, since shifting $B_{\rm a}$ at
constant $\nu$ means that we are sampling each time a different
portion of the NMR signal, which means different quadrupolar
satellites, different orientation of the crystallites, etc.
Nevertheless, the agreement with the estimate of $T_{1\text{n}}$
obtained by specific heat data (inset of Fig. \ref{CmvsT-B}) turns
out to be satisfactory. We can directly compare the NSLR rates
$1/T_{1 {\rm n}}(B_{\rm a})$ obtained from respectively NMR and
specific heat data by plotting (dashed line in Fig. \ref{W1W2Mn6})
$1/T_{1\text{n}}(B_{\rm a})$ as calculated from Eq. (\ref{T1Mn6}),
with $T=0.9$ K and fixing $\kappa_{0} = 40$~s$^{-1}$T$^{-3}$ as
obtained from the fit of $B^*(T^*)$ in the inset of Fig.
\ref{CmvsB}. The agreement is seen to be reasonable. The solid
line in the same figure represents a fit to the $1/T_{1 {\rm
n}}(B_{\rm a})$ based on a model described by Eq. (\ref{T1star})
of the following section, with parameters given in the discussion
there.

In Fig. \ref{W1W2Mn6} we also show the TSSR rate
$1/T_{2\text{n}}(B_{\rm a})$, with ordinate axis shifted in order
to compare its field dependence to that found for the NSLR rate.
As explained below, we expect $1/T_{2\text{n}}(B_{\rm a}) \propto
\exp(g \mu_{\rm B} B_{\rm a} / k_{\rm B} T)$ at high fields. This
dependence, and the quantitative result calculated according to
Eq. (\ref{F.13}), are shown as a dotted line in Fig. \ref{W1W2Mn6}
and yield the right order of magnitude of the observed
$T_{2\text{n}}\approx 0.1-1$ ms.

\section{Discussion: Magnetic hyperfine interactions and nuclear magnetic
relaxation}

In this section we apply theoretical results for magnetic
hyperfine interactions and nuclear magnetic relaxation in magnetic
insulating solids to interpret the data of the previous sections.
We recall that for such materials direct relaxation channels such
as quadrupolar interactions connecting nuclear spins to the
lattice become ineffective at low temperatures. Thus the electron
spins present have to serve as an intermediary between nuclei and
phonons in some way or the other and we have to consider the
interconnected spin dynamics of  both nuclear and electron spin
systems and their coupling to the lattice. As will be shown below,
for highly polarized  electron spin systems, such as Mn$_6$ in
high fields at low temperature, the relaxation behavior can be
adequately described in terms of theoretical models previously
developed in the field of dynamic nuclear polarization. These same
models should also provide a good  basis to describe the
relaxation in highly anisotropic single-molecule magnets like
Mn$_{12}$-ac and Fe$_{8}$ below their blocking temperatures, where
an extreme polarization of the electron spins is induced by the
crystal field.  Although several groups have described the
application to molecular magnets of some general theories of
nuclear
relaxation,\cite{lascialfari98PRL,salman02CM,goto03PRB,pilawa05PRB}
it appears that the role of the electronic dipolar coupling has so
far been overlooked. We therefore consider worthwhile to give a
detailed overview of the different ingredients needed to arrive at
a consistent picture describing the behavior we observe in both
the NMR and the specific heat of Mn$_6$, being confident it will
be quite useful for the other materials mentioned as well. We will
briefly come back to this point in the conclusion section, drawing
a comparison with our observations in Mn$_{12}$-ac. We start with
an evaluation of the relevant hyperfine interactions between the
cluster spins and the various nuclear spins in the Mn$_6$
molecular cluster. Quite generally,\cite{Slichter78,abragam61} the
hyperfine interaction hamiltonian of a nuclear moment $\mathbf{I}$
with the surrounding electron spins $\mathbf{s}_{\rm j}$ can be
written in the form of the bilinear coupling
\begin{equation}
{\cal H}_{\mathrm{hf}}  =  \sum_{\rm j} \mathbf{I}\cdot
\tilde{A}\cdot \mathbf{s}_{\rm j}
\end{equation}
Here the hyperfine interaction $\tilde{A}$ is a second rank tensor
and summation is over electron spins on both the same atom and
surrounding atoms. It is often convenient to interpret the
hyperfine coupling in terms of a magnetic hyperfine field: $B_{\rm
hf} = -(\gamma_{\rm n} \hbar)^{-1} \mathbf{s}\cdot\tilde{A}$,
acting on the nuclear spin in addition to the applied field
$B_{\rm a}$. $\gamma_{\rm n}$ is the nuclear gyromagnetic ratio.
The total magnetic Hamiltonian for the nuclear spins then becomes
\begin{equation}
{\cal H}_{\rm hf} = - \gamma_{\rm n} \hbar
\mathbf{I}\cdot(\mathbf{B}_{\rm a} + \mathbf{B}_{\rm hf}) =
-\gamma_{\rm n} \hbar \mathbf{I}\cdot \mathbf{B}_{\rm tot}
\end{equation}
In principle ${\cal H}_{\rm hf}$ will be time-dependent since both
$\tilde{A}$ and $\mathbf{s}_{\rm j}$ can depend on time due to,
respectively, atomic motions (not considered here) and
fluctuations of the electron spins. Provided that the frequency of
these fluctuations is fast compared to the nuclear Larmor
frequencies produced in the static case, the nuclear resonance is
still well defined be it at a frequency that is shifted with
respect to that for $\mathbf{B}_{\rm hf} = 0$, the shift being
proportional to the time average of of $\tilde{A}(t) \cdot
\mathbf{s}$. In the effective field picture the hyperfine field
can be split up into a static part, $\langle{\mathbf{B}}_{\rm
hf}\rangle$, and a time-dependent part $\mathbf{B}_{\rm hf}(t) =
\langle{\mathbf{B}}_{\rm hf}\rangle + \mathbf{b}(t)$. The
time-dependent fraction, $\mathbf{b}(t)$, is usually much smaller
and can then be treated as a perturbation that may produce
relaxation of the nuclear polarization. Neglecting the quadrupolar
interactions, the remaining magnetic hyperfine interaction may be
decomposed into contributions coming from the coupling of the
nuclear moment with the orbital motions of the electrons, the
dipolar interactions with the electron spins and, in case a finite
density of electrons is present at the nuclear site, the part due
to the Fermi-contact interaction. Thus we may write
$\langle{\mathbf{B}}_{\rm hf}\rangle  = \mathbf{B}_{\rm dip} +
\mathbf{B}_{\rm L} + \mathbf{B}_{\rm F}$, in an obvious notation.
As for the relative strengths of these contributions, for nuclei
(such as the present $^{55}$Mn) residing on magnetic atoms the
Fermi-contact term is strongest by far, with $B_{\rm F} \sim 10 -
10^{2}$~T, followed by the orbital and dipolar interactions with
electrons on the same atom (on site), typically of order 10 T. For
ions with closed or half-filled shells both $B_{\rm L}$ and
$B_{\rm dip}$ vanish, whereas also for non S-state ions, such as
Mn$^{3+}$, the orbital moment can be quenched by crystal-field
splittings, so that $B_{\rm L}$ becomes negligible. Furthermore,
nuclei on nonmagnetic ligand atoms directly coordinating the
magnetic ions via covalent bonds may also experience substantial
Fermi-contact interaction, of order $1-10$ T. Nuclei on more
distant nonmagnetic atoms will mainly experience the long-range
dipolar interactions, giving typically $B_{\rm dip} \sim 0.1-1$~T.

For the $^{55}$Mn nuclei on the Mn$^{3+}$ ions (with nuclear spin
$I=5/2$ and electron spin $s=2$) of the present compound Mn$_{6}$
we deduced in section III-E a resonance frequency (extrapolated to
$B_{\rm a} = 0$) centered around $360$ MHz (corresponding to
$\langle{B}_{\rm hf}\rangle = 34$ T), which is almost the same as
the value $364$ MHz Kubo \textit{et al.} \cite{kubo02PRB} recently
found for one of the two Mn$^{3+}$ sites in Mn$_{12}$-ac.
According to their analysis, the corresponding $B_{\rm hf} = 34.5$
T for this site results from the combination of Fermi-contact and
dipolar fields $B_{\rm F} = 41$ T and $B_{\rm dip} = 14$ T that
are of opposite sign. As they point out, the value for $B_{\rm F}$
should not vary much for the same ion in comparable coordinations,
quoting values for Mn$^{3+}$ in TiO$_{2}$ and in MnFe$_{2}$O$_{4}$
of, respectively, $B_{\rm F} = 42$ T and $36$ T (with $B_{\rm dip}
= 12$ T and $11.5$ T, again of opposite sign). We may therefore
assume similar values for these fields in Mn$_{6}$. As is often
done, in order to estimate the resulting nuclear energy level
splittings responsible for the measured hyperfine specific heat,
we have in sections III B and D approximated the net time average
of the diagonal part of the (slightly anisotropic
\cite{kubo02PRB}) hyperfine interaction by an effective isotropic
scalar interaction ${\cal H}_{\rm hf,is} = A
\mathbf{s}\cdot\mathbf{I}$, where $A = -\gamma_{\rm n} \hbar
B_{\rm hf} /s$, $s = 2$ and $A/k_{\rm B} = 8.7$ mK (corresponding
to $B_{\rm hf} = 34$ T). As for the other nuclei present in
Mn$_{6}$, the only sizable contribution to be expected is that
arising from the long-range dipolar interactions of the proton
spins with the surrounding electronic spins. Since these dipolar
fields are small, their contribution may be neglected in zero
applied field as compared to that of the $^{55}$Mn nuclei, and
only becomes substantial for applied fields exceeding $1$~T.
Although transferred hyperfine interactions with nuclei on the
oxygen and Br ions that are directly bonded covalently to the Mn
atoms could be substantial,\cite{kubo02PRB} we may nevertheless
neglect their contributions in comparison with the other ones in
view of the low abundance of the $^{17}$O isotope and the low
number of Br atoms present.

Considering next the nuclear spin relaxation (NSR) we remark that
the longitudinal NSR rate, $1/T_{\rm 1n}$, is given quite
generally by the expression
\begin{equation}
\frac{1}{T_{\rm 1n}} = \frac{1}{2} \sum_{m,n}
W_{m,n}(E_{m}-E_{n})^{2}/\sum_{m} E_{m}^{2}, \label{F.1}
\end{equation}
where $W_{m,n}$ denotes the probability for a transition between
nuclear energy levels $m,n$ induced by the perturbation
considered. As mentioned, we assume the main source for NSR to be
the time-dependent fluctuations of the electron spins
$\mathbf{s}(t)$ that produce fluctuating components
$\mathbf{b}(t)$ of the hyperfine field. The theory has been
developed by Moriya,\cite{moriya56PTP,moriya58JPSJ,moriya58JPSJb}
on basis of the general theory of magnetic resonance absorption of
Kubo and Tomita.\cite{Kubo54} Two possible sources for the
electron spin fluctuations have to be considered, namely electron
spin-lattice relaxation, characterized by the longitudinal
electron spin-lattice relaxation rate $1/T_{1 {\rm e}}$, and
spin-spin relaxation, due to magnetic interactions (dipolar or
exchange) between the electron spins and characterized by the
transverse electron spin relaxation rate $1/T_{2 {\rm e}}$.
Obviously, hyperfine interaction terms producing NSR should
involve operator combinations as $I^{\pm}s^{z}$, $I^{+} s^{-}$ and
$I^{-}s^{+}$, $I^{+}s^{+}$ and $I^{-}s^{-}$. The first of these
distinguishes itself in that a nuclear spin flip is \emph{not}
combined with an electronic spin flip. It follows that this first
type involves transitions at NMR frequencies $\omega=\omega_{\rm
n}$, the others at ESR frequencies $\omega = \omega_{\rm e} \pm
\omega_{\rm n} \approx \omega_{\rm e}$. In the high-$T$
approximation, the NSR rate can be expressed in terms of the
spectral densities $f_{\rm j}^{\alpha}$ of the two-spin ($i \neq
j$) and autocorrelation ($i=j$) functions $\langle
s_{i}^{\alpha}(0)s_{j}^{\alpha}(t) \rangle$ at these frequencies
as
\begin{eqnarray}
\frac{1}{T_{1 {\rm n}}} &=& \frac{2}{3} s(s+1) \sum_{j} \left[
\text{A}_{j} f_{j}^{z}(\omega_{\rm n}) + \text{B}_{j}
f_{j}^{\pm}(\omega_{\rm e}) \right], \quad \label{F.2}\\
f_{j}^{\alpha}(\omega) &=& \int_{-\infty}^{+\infty} \langle
s_{i}^{\alpha}(0) s_{j}^{\alpha}(t)\rangle e^{-i \omega t}
\mathrm{d}t \;(\alpha = \pm,z).
\end{eqnarray}
The coefficients $\text{A}_{j}$ and $\text{B}_{j}$ are constants
depending on the details of the hyperfine interactions. For the
ease of discussion we approximate the Fermi-contact interaction by
an isotropic scalar on-site hyperfine coupling and neglect
possible orbital contributions. Adding the dipolar interaction,
the hyperfine Hamiltonian becomes
\begin{eqnarray}
&{\cal H}_{\rm hf}& = A \mathbf{I}_{i}\cdot\mathbf{s}_{i}
\nonumber \\
&+& \frac{\mu_0}{4\pi}\gamma_{\rm e} \gamma_{\rm n} \hbar^2
\sum_{j} \left[\frac{\mathbf{I}_{i}\cdot
\mathbf{s}_{j}}{r_{ij}^{3}} - 3\frac{(\mathbf{I}_{i}\cdot
\mathbf{r}_{ij})(\mathbf{s}_{j}\cdot\mathbf{r}_{ij})}{r_{ij}^{5}}\right],
\qquad \label{F.3}
\end{eqnarray}
where $\gamma_{\rm e}$ is the electronic gyromagnetic ratio. For
the terms responsible for NSR one then obtains
\begin{eqnarray}
{\cal H}^{\prime} = && I_{i}^{+} \sum_{j} D_{ij}^{z} s_{j}^{z}(t)
\nonumber \\ &+& I_{i}^{+} \left[(1/2) A s_{i}^{-} + \sum_{j}
(D_{ij}^{-} s_{j}^{-} + D_{ij}^{+} s_{j}^{+})\right] \nonumber \\
&+& c.c., \label{F.4}
\end{eqnarray}
where $D_{ij}^{z}$ and $D_{ij}^{\pm}$ denote the components of the
dipolar coupling tensor connecting $I_{i}^{+}$ with $s_{j}^{z}$
and with $s_{j}^{\pm}$, respectively, and $c.c.$ stands for
complex conjugates. It is important to note that for a pure scalar
hyperfine interaction only the transverse spectral densities
$f_{j}^{\pm} (\omega_{\rm e})$ do appear. Assuming an exponential
decay of the spin-correlation functions, the scalar interaction
leads to the NSR rate\cite{abragam61}
\begin{eqnarray}
\frac{1}{T_{1 \rm{n}}} &=& \frac{1}{2} A^{2}
\int_{-\infty}^{+\infty} \langle s^{+}(0)s^{-}(t) \rangle e^{\-i
(\omega_{\rm e} - \omega_{\rm n}) t} dt \nonumber \\ &=&
\frac{1}{3} s(s+1)A^{2} \frac{T_{2 \rm e}}{1+(\omega_{\rm
e}-\omega_{\rm n})^{2}T_{2 \rm{e}}^{2}}. \label{F.5}
\end{eqnarray}
As noted above, relaxation then requires energies $\hbar
\omega_{\rm e}$ of the order of the electronic level splittings.
By contrast the dipolar interaction contains terms of different
symmetry, \cite{abragam61,Slichter78} so that it contributes to
both transverse and longitudinal terms. In particular it contains
the operator $D_{ij}^{z} \propto -(3/2) \sin \theta \cos \theta
e^{-i \phi}$ that may induce a nuclear flip unaccompanied by an
electron flip, thus involving the much smaller energy $\hbar
\omega_{\rm n}$. The NSR rate due to this process, after averaging
over the angular dependence, is obtained as\cite{abragam61}
\begin{equation}
\frac{1}{T_{1 {\rm n}}} =
\frac{3}{5}\left(\frac{\mu_0}{4\pi}\right)^2(\gamma_{\rm e}
\gamma_{\rm n} \hbar)^{2} r^{-6} \int_{-\infty}^{+\infty} \langle
s^{z}(0) s^{z}(t) \rangle e^{-i \omega_{\rm n} t} dt.
\end{equation}

Considering now first the case that the electron spin fluctuations
arise from electron spin-lattice relaxation, and assuming again an
exponential decay of the autocorrelation function, in this case
with longitudinal relaxation rate  $1/T_{1 {\rm e}}$:
\begin{equation}
\langle s^{z}(0) s^{z}(t) \rangle = \frac{1}{3} s(s + 1) \exp(-t/
T_{1 {\rm e}}), \label{F.6}
\end{equation}
the NSR rate by this process is found to be given by
\begin{equation}
\frac{1}{T_{1 {\rm n}}^{\rm (EZ)}} = \frac{2}{5}
\left(\frac{\mu_0}{4\pi}\right)^2 (\gamma_{\rm e} \gamma_{\rm n}
\hbar)^{2} r^{-6} s(s+1) \frac{T_{1 {\rm e}}}{1+\omega_{\rm n}^{2}
T_{1 {\rm e}}^{2}}, \label{F.7}
\end{equation}
an expression first derived by Bloembergen\cite{Bloembergen49} for
the nuclear spin-lattice relaxation by paramagnetic impurities in
diamagnetic crystals. The superscript (EZ) is added to indicate
that this nuclear relaxation is driven by the spin-lattice
relaxation of the Electron-Zeeman reservoir. Comparing Eqs.
(\ref{F.5}) and (\ref{F.7}) it is clear that, unless the
electronic linewidths would be comparable to the level splittings
$\omega_{\rm e}$, the latter process will outweigh the previous
one by the large factor $(\omega_{\rm e}/\omega_{\rm n})^{2}$. In
most cases of interest one further has $\omega_{\rm n} T_{1 {\rm
e}} \gg 1$, so that one may write approximately
\begin{eqnarray}
\frac{1}{T_{1 {\rm n}}^{\rm (EZ)}} &\approx& \frac{2}{5}
\left(\frac{\mu_0}{4\pi}\right)^2 \frac{(\gamma_{\rm e}
\gamma_{\rm n} \hbar)^{2} r^{-6} s(s + 1)}{\omega_{\rm n}^{2} T_{1
{\rm e}}} \nonumber \\ &=& \frac{2}{5} \left( \frac{B_{\rm
dip}}{B_{\rm tot}}\right)^{2} \frac{1}{T_{1 {\rm e}}} \label{F.8}
\end{eqnarray}
Here $B_{\rm dip} = (\mu_0/4\pi)\hbar \gamma_{\rm e}
\sqrt{s(s+1)}r^{-3}$ stands for the electronic dipolar field at
the nuclear site, and $B_{\rm tot} = B_{\rm a} + B_{\rm hf} =
\omega_{\rm n} / \gamma_{\rm n}$ is the total field responsible
for the nuclear Zeeman splittings.\cite{abragam61}

In the next step we have to compare this result with the NSR rate
arising from spin-spin interactions, which for Mn$_{6}$ amount to
the dipolar interactions between the electronic cluster spins
$S=12$. We note that since the total spin $S=12$ of the Mn$_{6}$
cluster results from the strong ferromagnetic intramolecular
exchange between the atomic spins $s=2$, the fluctuations of the
total spin are obviously related to those of the constituting
atomic spins and vice versa. Thus, although both the hyperfine
interactions and the electron spin lattice coupling basically
involve the atomic spins, the atomic spin fluctuations
nevertheless are in a one-to-one relationship with those of the
cluster spins. Due to such spin-spin interactions the spectral
density will no longer be given by a Lorentzian. Instead of the
exponential decay of the correlation functions, Eq. (\ref{F.6}),
one usually assumes a Gaussian approximation for the
autocorrelation
functions\cite{moriya56PTP,moriya58JPSJ,moriya58JPSJb}:
\begin{eqnarray}
\langle s^{z}(0) s^{z}(t) \rangle &=&  \frac{1}{2} \langle
s^{+}(0) s^{-}(t)\rangle \nonumber \\
&=& \frac{1}{3} s(s+1) \exp(-\omega_{\rm int}^{2} t^{2}).
\label{F.9}
\end{eqnarray}
For the longitudinal NSR rate one obtains:
\begin{equation}
\frac{1}{T_{1 {\rm n}}^{\rm (ED)}} = \frac{\sqrt{2 \pi}}{3}
\gamma_{\rm n}^{2} B_{\rm dip}^{2} \omega_{\rm int}^{-1} \exp(-
\omega_{\rm n}^{2}/2 \omega_{\rm int}^{2}), \label{F.10}
\end{equation}
where (ED) indicates that this process is driven by fluctuations
in the Electron-Dipolar reservoir. For the transverse relaxation
one finds similarly:
\begin{equation}
\frac{1}{T_{2 {\rm n}}} = \frac{1}{2T_{1 {\rm n}}^{\rm (ED)}} [1 +
\exp(- \omega_{\rm n}^{2}/2 \omega_{\rm int}^{2})], \label{F.11}
\end{equation}
from which it follows that $1/T_{2 {\rm n}} \approx 1/T_{1 {\rm
n}}^{\rm (ED)}$. Here $\omega_{\rm int}$ stands for the electronic
dipolar spin-spin interaction,\cite{Abragam82} which in our case
can be estimated from the dipolar ordering temperature, $\hbar
\omega_{\rm int} \approx k_{\rm B} T_{\rm c}$, and also
corresponds to the electronic TSSR rate, $\omega_{\rm int} \approx
1/T_{2 {\rm e}}$.

At this point it is important to emphasize that the above
derivations are essentially only valid at high temperatures and
low applied fields, since the effects of polarization of the
electronic spins by the applied field have been neglected. As
noted already by Moriya\cite{moriya58JPSJ} and in later work on
dynamic polarization,\cite{Gunter67,Lowe68} the more the electron
spins become polarized, the less they will be able to relax the
nuclear spins. To account for this, one should replace the time
dependencies of the electronic spin $\mathbf{s}(t)$ by its
fluctuating part, $\delta \mathbf{s}(t) = \mathbf{s}(t) -
\mathbf{s}_{0}$, where $\mathbf{s}_{0}$ denotes the thermal
average of $\mathbf{s}$. Thus, instead of an expression as in Eq.
(\ref{F.6}) for the decay of the electronic spin, one should
take\cite{moriya58JPSJ}
\begin{eqnarray}
\langle \delta s^{z}(0) \delta s^{z}(t) \rangle &=& \langle
(s^{z}(0) - s^{z}_{0}) (s^{z}(t) - s^{z}_{0}) \rangle \nonumber \\
&=& \langle (s^{z} - s^{z}_{0})^{2} \rangle \exp(-t/T_{1 {\rm e}}) \nonumber \\
&=& S (\partial/ \partial X) s^{z}_{0}(X) \exp(-t/T_{1 {\rm e}})
\nonumber \\ &=& S s (\partial/\partial X) B_{S}(X) \exp(-t/T_{1
{\rm e}}), \qquad
\end{eqnarray}
using the fact that the thermal average of each Mn$^{3+}$ spin
$s^{z}$ is given by $sB_{S}(X)$, where $X = g \mu_{\rm B}B_{\rm
a}S/k_{\rm B}T$, $s=2$, and $B_{S}$ is the Brillouin function for
the {\em total} molecular spin $S$. Accordingly, the expression
(\ref{F.6}) for the electron spin autocorrelation function should
be multiplied by the factor $3S(s + 1)^{-1}
\partial B_{\rm S}/\partial X$. Restricting in what follows to
the simplest case of spin $S = 1/2$, as appropriate for the
present experiments in the high-field / low-$T$ range where only
the two lowest lying electron Zeeman states are relevant, this
factor reduces to $(1 - \tanh^{2}X)/2$, with $X = g \mu_{\rm
B}B_{\rm a}/2k_{\rm B}T$. We thus obtain for $1/T_{1 {\rm n}}^{\rm
(EZ)}$ instead of Eq. (\ref{F.8}) the relation
\begin{equation}
\frac{1}{T_{1 {\rm n}}^{\rm (EZ)}} \approx \frac{1}{5} \left(
\frac{B_{\rm{dip}}}{B_{\rm tot}} \right)^{2} (1 - \tanh^{2} X)
\frac{1}{T_{1 {\rm e}}} \label{F.12}
\end{equation}
whereas instead of Eq. (\ref{F.10}) one has now:
\begin{eqnarray}
\frac{1}{T_{1 {\rm n}}^{\rm (ED)}} \approx \frac{1}{T_{2 {\rm
n}}^{\rm (ED)}} \qquad \qquad \qquad \qquad \nonumber\\
 \approx \frac{\sqrt{2\pi}}{6}\frac{(\gamma_{\rm n} B_{\rm
dip})^{2}}{ \omega_{\rm int}} (1 - \tanh^{2}X) \exp(-\omega_{\rm
n}^{2}/2 \omega_{\rm int}^{2}). \quad \label{F.13}
\end{eqnarray}
In  both cases, since $\tanh X$ gives the degree of polarization
of the electron spin, one observes that when this approaches unity
the nuclear relaxation rate goes to zero, as to be expected. For
the electron-dipolar relaxation channel one should notice that,
although the actual electronic linewidth $1/T_{\rm 2e}$ strongly
depends on the electronic polarization (the second moment of the
absorption line is proportional to $1 - \tanh^{2}X$, cf. Ref.
\onlinecite{abragam70}), $\omega_{\rm int}$ in (\ref{F.13}) is
still given by the dipolar coupling as calculated in the high-$T$
limit.\cite{Abragam82}

Proceeding next to compare the above predictions with the
high-field NMR experiment, we may already notice that Eq.
(\ref{F.13}) yields the right order of magnitude for $1/T_{2 {\rm
n}}$. From the value of $T_{\rm c} \approx 0.16$ K, we deduce the
electronic dipolar broadening to be $\omega_{\rm int} \approx 2
\times 10^{10}$ rad/s. With NMR frequencies of order $\omega_{\rm
n} \approx 1.5 \times 10^{9}$ rad/s the factor $\exp(-\omega_{\rm
n}^{2}/2\omega_{\rm int}^{2})$ becomes $\approx 1$. Further, we
have $B_{\rm tot} \approx 30$ T and $B_{\rm dip} \approx (1/3)
B_{\rm tot} \approx 10$ T, yielding $\gamma_{\rm n} B_{\rm dip}
\approx 6 \times 10^8$ rad/s. For applied fields $B_{\rm a}
> 5$ T the polarization correction factor $(1 - \tanh^{2}X)$
becomes of order $10^{-3}$ to $10^{-4}$. From Eq. (\ref{F.13})
with the numerical factors quoted above we thus find the
prediction (cf. Fig. \ref{W1W2Mn6}, dotted line): $1/T_{2 {\rm
n}}^{\rm (ED)} \approx 10^{4}$ to $10^{3}$ s$^{-1}$ for the
transverse NSR-rate arising from electron spin-spin interactions,
i.e. in the same range as the experimental transverse rate.
Conversely, the data clearly show that $1/T_{1 {\rm n}} \ll 1/T_{2
{\rm n}}$, contrary to the prediction of Eq. (\ref{F.13}). Indeed,
this process basically only establishes the thermal equilibrium
between the nuclear and the electronic spin systems, i.e. without
considering the relaxation of the latter toward the phonon bath.
For the complete description of the nuclear spin-lattice
relaxation process we obviously have to investigate the
spin-phonon coupling mechanism as well.

In order to estimate the electronic $1/T_{1 {\rm e}}$, we remark
that the electron spin-lattice relaxation rate arising from
transitions between the two lowest Zeeman levels of the $S = 12$
multiplet due to the direct process will be given by the sum of
the transition rates $w_{\uparrow}$ and $w_{\downarrow}$ due to
absorption and emission of phonons, respectively\cite{abragam70}
\begin{equation}
\frac{1}{T_{1 {\rm e}}} = w_{\uparrow} + w_{\downarrow}.
\label{T1e}
\end{equation}
Since the phonon modulation of the crystal field can be expected
to be the main source of this coupling, we may apply to Mn$_6$ the
calculations developed by Leuenberger and Loss,
\cite{leuenberger00PRB} obtaining
\begin{subequations}
\begin{eqnarray}
w_{\uparrow} &=& V_{\rm e-ph}(g\mu_{\rm B}B_{\rm a})^3
\frac{1}{\exp(2X) - 1},\\
w_{\downarrow} &=& V_{\rm e-ph}(g\mu_{\rm B}B_{\rm a})^3
\frac{1}{1-\exp(-2X)},\\
V_{\rm e-ph} &=& \frac{D^2 S(2S-1)^2}{6\pi \rho c_{\rm s}^5
\hbar^4}, \label{Veph}
\end{eqnarray}
\end{subequations}
where $S=12$, $D/k_{\rm B} = 0.013$ K is the uniaxial anisotropy
constant, $\rho = 1.45$ g/cm$^{3}$ is the density and $c_{\rm s}$
the sound velocity. Within the Debye model, the latter is obtained
from the experimental Debye temperature $\Theta_{\rm D} = 29$ K as
\begin{equation}
c_{\rm s} = \frac{k_{\rm B} \Theta_{\rm D}}{\hbar}\left( \frac{6
\pi N_{\rm A}}{V_{\rm m}} \right)^{-1/3} = 1.3 \times 10^{3} \;
\mathrm{m}/\mathrm{s}.
\end{equation}
Substituting into Eq. (\ref{T1e}) yields
\begin{equation}
1/T_{1{\rm e}} \approx 104 \; B_{\rm a}^{3} \coth X,
\label{T1efinal}
\end{equation}
with $B_{\rm a}$ in Tesla. (It should be noted that the value
calculated for $V_{\rm e-ph}$ is very sensitive to the values used
for $\Theta_{\rm D}$ and $D/k_{\rm B}$ so that it obviously is
subject to a large uncertainty margin). For instance, $B_{\rm a} =
5$ T and $T = 1$ K yields $1/T_{\rm 1e} \sim 10^4$ s$^{-1}$.
Because of the very small value of the anisotropy constant $D$ in
Mn$_6$, $1/T_{\rm 1e}$ is thus expected to be much lower than the
typical values $\sim 10^7$ s$^{-1}$ found, e.g., in Mn$_{12}$-ac.
This also implies that a model for the TSSR rate based on the
random changes in local hyperfine field due to electron-phonon
excitations, as recently used to describe $1/T_{\rm 2n}$ in
Mn$_{12}$-ac,\cite{goto03PRB} would lead in this case to a
quantitative estimate that is about three orders of magnitude
lower than our experimental result.

Nuclear relaxation to the lattice can now occur in two ways,
either directly via the spin-lattice relaxation fluctuations of
the individual electron spins, or in a two-step process by
spin-spin relaxation to the electron dipolar reservoir followed by
relaxation to the lattice. The direct spin-lattice relaxation
(single-ion) process is described by Eq. (\ref{F.12}), which
becomes
\begin{equation}
\frac{1}{T_{1 {\rm n}}^{\rm (EZ)}} \approx 21 \left(\frac{B_{\rm
dip}}{B_{\rm tot}}\right)^{2} B_{\rm a}^{3} \coth X (1 - \tanh^{2}
X). \label{F.14}
\end{equation}
For large $X$, $\coth X \approx 1$ and $(1-\tanh^{2} X) \approx 4
\exp(-2X)$. With $B_{\rm dip}/B_{\rm tot} \approx 1/3$ one obtains
\begin{equation}
\frac{1}{T_{1 {\rm n}}^{\rm (EZ)}} \approx 9 \; B_{\rm a}^{3}
\exp(-g \mu_{\rm B}B_{\rm a}/k_{\rm B}T), \label{F.15}
\end{equation}
(with $B_{\rm a}$ in Tesla). As we have seen, both the specific
heat and the NMR data yield a field dependence of $1/T_{1 {\rm
n}}$ that is the same as in Eq. (\ref{F.15}), with a prefactor of
about 40. The prediction of Eq. (\ref{F.15}) is therefore
qualitatively satisfactory but quantitatively slightly too low.

Next we consider the two-step relaxation process on basis of the
spin-spin interaction process. Intuitively this is easily
understood as follows: relaxation of the nuclei by spin-spin
interactions involves $1/T_{2 {\rm e}}$ which will be of order
$10^{9}$ Hz or higher, implying that the electron spin-spin
interactions can be very effective in relaxing the nuclear spins.
However, relaxation is then toward the electron spin system, and
the ultimate relaxation to the lattice has to occur in a second
step. This situation has often been met for nuclear relaxation in
magnetic crystals or in diamagnetic insulators with paramagnetic
impurities, notably  in connection with the phenomenon of dynamic
nuclear polarization.\cite{Abragam82,Abragam80,Khutsishvili70} In
the theoretical treatments it has been proven necessary to
consider the Zeeman term and the spin-spin interaction term in the
Hamiltonian of a spin system (electronic or nuclear) as separate
energy reservoirs, to each of which separate temperatures can be
assigned that may differ quite substantially from one another (see
Refs. \onlinecite{Abragam82} and \onlinecite{Khutsishvili70}).

\begin{figure}[t]
\includegraphics[width=7.5cm]{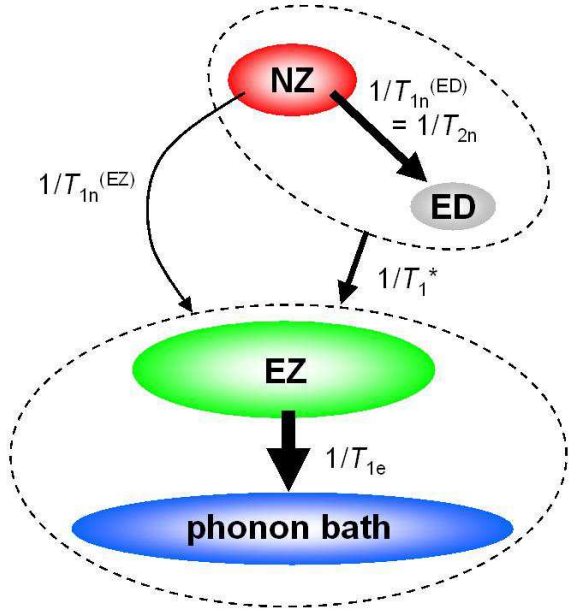}
\caption{\label{blockdiagram} (Color online) Block diagram of the
nuclear and electronic spin systems involved in the relaxation
process at high fields, and the relative rates of energy
transfer.}
\end{figure}

Applied to our present problem, this leads to the block diagram
sketched in Fig. \ref{blockdiagram} (the Nuclear-Dipolar reservoir
is omitted here since its energy is so small that it plays no role
at the relevant temperatures). The Electron-Zeeman (EZ) and
Electron-Dipolar (ED) energy reservoirs will be at the same
(lattice) temperature in zero applied field. However, when with
increasing field the electronic level splitting $\omega_{\rm e}$
starts to exceed the electronic dipolar broadening, the two
reservoirs  become progressively separate entities, characterized
by different temperatures and largely different heat capacities.
This arises since the EZ reservoir is strongly coupled to the
phonon bath and can be considered to remain in equilibrium with
the lattice regardless of any nuclear relaxation event. The ED
reservoir on the other hand, has a heat capacity that decreases
rapidly with field. Although the ED reservoir is coupled to the
lattice at a rate $\approx 2/T_{1 {\rm e}}$, i.e., twice the
electron spin-lattice relaxation rate,\cite{Khutsishvili70} it is
also strongly coupled to the nuclear Zeeman (NZ) system at a rate
given by Eq. (\ref{F.13}), i.e., of the order of $1/T_{2 {\rm
n}}$. In high fields, therefore, the nuclear relaxation will take
place in two stages. In the first stage the NZ and ED systems will
become rapidly in equilibrium at basically the same fast rate that
determines the experimentally observed $1/T_{2 {\rm n}}$.
Subsequently, the coupled NZ + ED systems will relax toward the
lattice at the much slower rate
\begin{equation}
\frac{1}{T_{1}^{\ast}} = \frac{2}{T_{1 {\rm e}}} \frac{C_{\rm
ED}}{C_{\rm NZ}+C_{\rm ED}},
\end{equation}
which for the case $C_{\rm NZ}\gg C_{\rm ED}$ would become roughly
equal to $2/T_{1 {\rm e}}$ ($C_{\rm ED}/C_{\rm NZ})$. Here the
symbols $C_{\rm ED}$ and $C_{\rm NZ}$ stand for the
(field-dependent) specific heats of the ED and the NZ reservoirs.
The situation is seen to be analogous to the phonon-bottleneck
phenomenon, well known in paramagnetic relaxation. Applied to our
present case, we calculate $C_{\rm NZ}$ from Eq. \ref{cnucleq} to
be of order 0.006 R at T = 0.9 K and B = 5 T (cf. Figs. 6 and 7),
which value depends only weakly on applied field. $C_{\rm ED}$ in
zero field can be estimated as:\cite{abragam70}
\begin{eqnarray}
\frac{C_{\rm ED}}{R} &=& \frac{6}{5} \left( \frac{\mu_0}{4 \pi}
\frac{g^2 \mu_{\rm B}^2 S(S+1)}{3k_{\rm B}T} \right)^2 \sum_{j>i}
\frac{1}{r_{ij}^6},
\end{eqnarray}
after averaging over the angular dependence of the dipolar
coupling, as appropriate for an unoriented powder sample. Using
the same lattice parameters as for the Monte Carlo simulations
yields $C_{\rm ED} \approx 0.004$ R at $T = 0.9$~K. $C_{\rm ED}$
then depends on the electronic polarization as $C_{\rm ED}(X) =
C_{\rm ED}(0)(1 - \tanh^2 X) \approx 4 \: C_{\rm ED}(0) \exp(-g
\mu_{\rm B} B_{\rm a} / k_{\rm B}T)$, since for high fields only
the lowest Zeeman level is available for the electron spins. All
this leads to a global rate for the nuclear spin-lattice
relaxation
\begin{eqnarray}
\frac{1}{T_1^{\ast}} & \approx & 208 \; B_{\rm a}^3 \coth(X)
\frac{0.004 (1 - \tanh^2 X)}{0.006 + 0.004 (1 - \tanh^2 X)}
\nonumber
\\
& \approx & 550 \; B_{\rm a}^3 \exp(-g \mu_{\rm B} B / k_{\rm B}
T), \label{T1star}
\end{eqnarray}
that has the same field dependence as Eq. (\ref{F.15}), but with a
prefactor of about 550 instead of 9, meaning that the two-step
spin-spin relaxation should be the fastest process by almost two
orders of magnitude in the high-field region. The solid line in
Fig. \ref{W1W2Mn6} is obtained from Eq. (\ref{T1star}) but
assuming a prefactor of order 25.

At this point it is important to recall that the prefactors in
both Eqs. (\ref{F.15}) and (\ref{T1star}) are affected by a large
numerical uncertainty originating from the expression for
$1/T_{\rm 1e}$, which contains the electron-phonon coupling
constant $V_{\rm e-ph}$, Eq. (\ref{Veph}). This constant is
proportional to $D^2$ and $\Theta_{\rm D}^{-5}$, and both these
quantities have rather large error bars. However, the values of
$D$ and $\Theta_{\rm D}$ influence \emph{in the same way} the
relaxation rates via the Electron-Zeeman and the Electron-Dipolar
channels, thus the latter is expected to dominate \emph{in any
case} by almost two orders of magnitude. Due to their strong
influence on $1/T_{\rm 1e}$, allowing both $D$ and $\Theta_{\rm
D}$ to vary by only a factor 1.5 would already yield the correct
quantitative prefactor in Eq. (\ref{T1star}).

Summarizing the results of this section, we may state that both
the longitudinal and transverse nuclear relaxation that we observe
at high applied fields are in excellent qualitative and even
quantitative agreement with the model based on fast dipolar
relaxation of the hyperfine-coupled Nuclear-Zeeman system to the
Electron-Dipolar reservoir, followed by much slower relaxation of
the combined systems via the electron spin-lattice channel. The
direct nuclear spin-lattice relaxation process by single-ion
electron spin-lattice relaxation predicts a similar field
dependence but is calculated to be much slower at high fields. At
low fields one will have $\omega_{\rm e} \approx \omega_{\rm
int}$, so that the Electron-Dipolar and the Electron-Zeeman
systems will become ``on speaking terms'', and a subdivision of
the two electron spin reservoirs is no longer valid. In this
range, however, NSR by means of the scalar hyperfine interaction,
Eq. (\ref{F.5}), should also become important (since then no
longer $\omega_{\rm n} \ll \omega_{\rm e}$, whereas $T_{2 {\rm e}}
\ll T_{1 {\rm e}}$).

\section{Concluding remarks}

In conclusion, our experiments on Mn$_{6}$ show that dipole-dipole
interactions between molecular magnetic clusters may indeed induce
long-range magnetic order at low temperatures if the anisotropy is
sufficiently small. Spin-lattice relaxation is then fast enough to
produce equilibrium conditions down to the low temperatures
needed. We should add that similar conditions could in principle
also be reached in the highly anisotropic cluster systems, for
which it was shown that by applying magnetic fields perpendicular
to the anisotropy axis, the spin-lattice relaxation can be tuned
and made similarly fast through the process of magnetic quantum
tunneling. However, it is very rare to observe magnetic ordering
phenomena in those systems. One exception known to
date\cite{evangelisti04PRL} originates from an unusually high
tunneling rate already in zero field. Instead, for the majority of
the anisotropic clusters it is likely that, given the magnitude of
the fields needed to have a considerable increase of the
relaxation rate ($B_{\perp} \gg 1$ T), any longitudinal component
of the field would create a Zeeman splitting that is much larger
than the energy involved in the magnetic dipolar ordering. We
found indeed that in Mn$_6$ the ordering transition is removed
already for relatively small fields ($\sim 0.5$ T). In a recent
neutron diffraction experiment on a Mn$_{12}$-ac single crystal,
however, Luis \textit{et al.}\cite{luis05PRL} achieved an
extremely accurate alignment of the field (to within 0.1 degree)
and obtained evidence for a ferromagnetic phase induced by the
transverse field.

We have also studied the nuclear spin dynamics of Mn$_6$, both
directly by NMR experiments and through the hyperfine contribution
to the field-dependent specific heat. The agreement between the
two techniques is very good, and also provides an interesting
comparison with the nuclear spin dynamics in the anisotropic
single-molecule magnet Mn$_{12}$-ac. Both qualitatively and
quantitatively, the nuclear magnetic relaxation data turn out to
be in good agreement with predictions obtained from theories
developed earlier for relaxation in paramagnetic crystals and for
dynamic polarization. In high fields, the observed nuclear
relaxation is dominated by electron spin fluctuations arising from
the dipolar interactions between cluster spins. In spite of the
large Zeeman splittings between the cluster spin levels produced
in such high fields, these fluctuations are able to relax the
nuclei through the dipolar part of the hyperfine interaction. In
this field range the electron dipolar and the electron Zeeman
system are basically decoupled. Relaxation of the nuclear spins
then proceeds in two steps, namely an initial rapid relaxation to
the electron dipolar system via the electron spin-spin interaction
channel, followed by a much slower relaxation of the combined
nuclear-electron spin systems to the lattice through the electron
spin-lattice channel. It is of interest in this regard to note
that the values for the longitudinal and transverse relaxation
rates as observed for Mn$_{12}$-ac in the low-temperature
($T<0.9$~K) quantum regime,\cite{morello04PRL} where the electron
spin fluctuations can be attributed to quantum tunnelling of the
cluster spins, fall only slightly below the present observations
for Mn$_{6}$ at $T=0.9$~K. Below the blocking temperature ($\sim
3$~K) the cluster spins of Mn$_{12}$-ac become almost fully
polarized even in zero field due to the strong crystal field
splittings of the electron spin levels, the distance between the
first excited state from the ground state amounting to more than
10 K. The temperature independent value found for the transverse
nuclear relaxation rate, $1/T_{2 {\rm n}} \approx 100$ s$^{-1}$,
could be well explained in terms of intercluster nuclear spin
diffusion, i.e., nuclear flip-flops arising from the dipolar
interaction between nuclear spins in neighboring clusters. The
same physical mechanism should also put a lower bound to the
transverse nuclear rate in Mn$_{6}$, which is, however, not
relevant due to the presence of the faster spin-spin relaxation
process.

For the longitudinal rate $1/T_{1 {\rm n}}$ for Mn$_{12}$-ac a
value of $\approx 0.03$ s$^{-1}$ is found below 1 K, slightly
depending on temperature. This value agrees with the
nuclear-spin-mediated tunneling rate estimated for the
fast-relaxing molecular spins in Mn$_{12}$-ac. As argued by
Morello \textit{et al}.,\cite{morello04PRL} the tunneling process
can at the same time provide a relaxation channel for the nuclei
to the electron-dipolar system. Similar to the above-discussed
case of Mn$_{6}$ in high field, relaxation to the lattice should
then occur in a second step through the electron spin-lattice
coupling.

\begin{acknowledgements} We thank J. Krzystek for performing
the high-field EPR study. This work is part of the research
program of the ``Stichting voor Fundamenteel Onderzoek der
Materie'' (FOM) and is partially supported by the European
Community contract no. MRTN-CT-2003-504880 ``QUEMOLNA'' and by the
EC-Network of Excellence ``MAGMANet'' (No. 515767-2). F. L.
acknowledges Grant Nº. PIP039/2005 from DGA. J. F. F. acknowledges
grant BFM2003-03919/FISI, from the MCyT of Spain.
\end{acknowledgements}

\ \\

\noindent $^*$ Corresponding author. Present address:\\
Department of Physics and Astronomy, University of British
Columbia, 6224 Agricultural Rd., Vancouver B.C. V6T 1Z1, Canada.\\
Electronic address: morello@physics.ubc.ca

\appendix

\section{Estimation of the effective demagnetizing factor}

In this Appendix, we derive an approximate expression for the
effective demagnetizing factor $N_{\rm eff}$ appropriate for the
cylindrically shaped container filled with the grains. This
problem is notoriously complex and can only be approximately
solved. In a first step the relation should be found between the
applied field $H_{\mathrm{a}}$ and the local field
$H_{\mathrm{loc}}$ acting on a reference grain in the container.
Approximating the grains as point dipoles, the difference
($H_{\mathrm{loc}} - H_{\mathrm{a}}$) will be due to the
contributions to the field arising from all the other dipoles
inside the container. Adopting the well-known Lorentz
construction, the dipole summation is split into one inside a
(sufficiently large) sphere around the reference grain and a
contribution from the dipoles outside this sphere. For this second
contribution the dipoles are usually assumed to form a homogeneous
continuum so that it is just proportional to the difference in
demagnetizing factors of the container ($N_{\mathrm{cont}}$) and
of the sphere ($1/3$). As for the first summation, it would be
zero for a cubic arrangement of the dipoles. This will not be the
case here since the grains are randomly packed, but since a valid
estimate is not easily obtained, and we may expect it to be small,
we shall just neglect it. One then obtains

\begin{equation}
H_{\mathrm{loc}}  =  H_{\mathrm{a}}  -  f M (N_{\mathrm{cont}}  -
1/3) \label{Hlocal}
\end{equation}
Here $f$ denotes the filling volume fraction of the grains in the
container and $M$ is the magnetization of the grains.

In the next step we have to correct $H_{\mathrm{loc}}$ for the
dipolar contributions arising from the magnetic material inside
the grain. In case of a ferromagnetic material, one usually only
takes the shape-dependent demagnetizing correction into account.
An argument for this may be found in that the magnetization
process for the ferromagnet is mostly determined by the mobility
of the domain walls, which will react to the macroscopically
averaged internal field. For simplicity, we first consider the
case of zero magnetocrystalline anisotropy for which the
demagnetization factor as well as the magnetization and fields can
be treated as scalars. We thus obtain for the internal field
$H_{\mathrm{i}}$ inside the grain

\begin{eqnarray}
H_{\mathrm{i}} & = &  H_{\mathrm{loc}}  -  N_{\mathrm{grain}} M
\nonumber \\ & = & H_{\rm a} - N_{\mathrm{grain}} M    -  f M (
N_{\mathrm{cont}} - 1/3 ) \label{Hinternal}
\end{eqnarray}

From the definition: $H_{\mathrm{i}}  =  H_{\mathrm{a}} -
N_{\mathrm{eff}}  M$, we thus finally find

\begin{equation}
N_{\mathrm{eff}}  = N_{\mathrm{grain}}  + f ( N_{\mathrm{cont}} -
1/3 ). \label{Neffiso}
\end{equation}

If, by contrast, the grain has uniaxial magnetic anisotropy, we
should distinguish between the parallel $\chi_{{\rm i,}\parallel}$
and the perpendicular $\chi_{{\rm i,}\perp}$ intrinsic
susceptibilities in the response to the internal magnetic field.
These susceptibilities depend on the magnitude of the anisotropy
parameter $D$. They can be obtained, from the numerically
calculated eigenstates and eigenvalues of the spin Hamiltonian,
using the Van Vleck's formalism as described in, e.g., Ref.
\onlinecite{Carlin86}. In this case Eq. (\ref{Hinternal}) becomes

\begin{equation}
\mathbf{H}_{\mathrm{i}}  = \mathbf{H}_{\mathrm{loc}}  -
\widetilde{N}_{\mathrm{grain}} \widetilde{\chi_{\rm i}}
\mathbf{H}_{\mathrm{i}}, \label{Hinternalanis}
\end{equation}
\noindent where $\widetilde{N}_{\mathrm{grain}}$ and
$\widetilde{\chi_{\rm i}}$ are respectively the diagonal
demagnetizing and intrinsic susceptibility tensors. By combining
Eqs.(\ref{Hlocal}) and (\ref{Hinternalanis}) it is possible to
find a relationship between the measured susceptibility $\chi$ and
the two components of $\widetilde{\chi_{\rm i}}$. For the case
when the anisotropy axes are randomly oriented in the sample, we
find

\begin{equation}
\chi = \frac{\chi_{\rm eff}}{1+f \chi_{\rm eff} (N_{\mathrm{cont}}
- 1/3)} \label{suscN1}
\end{equation}
\noindent where the susceptibility $\chi_{\rm eff}$ corrected for
the demagnetizing factor of the grains equals
\begin{equation}
\chi_{\rm eff} = \left[ \frac{2 \chi_{{\rm i,}\perp}}{3
\left(1+\chi_{{\rm i,}\perp}/3 \right)} + \frac{\chi_{{\rm
i,}\parallel}}{3 \left(1+\chi_{{\rm i},\parallel}/3 \right)}
\right]. \label{suscN2}
\end{equation}
This relationship was used to calculate the theoretical powder
susceptibilities shown in Fig. \ref{X12vsT}.


\begin{thebibliography}{99}


\bibitem{guntherB}
{\textit{Quantum Tunneling of the Magnetisation}, edited by L.
Gunther and B. Barbara, Kluwer (Dordrecht, 1995).}

\bibitem{chudnovskyB}
{E. M. Chudnovsky and J. Tejada, \textit{Macroscopic Quantum
Tunneling of the Magnetic Moment}, Cambridge University Press
(Cambridge, 1998).}

\bibitem{gatteschi02B}
{D. Gatteschi and R. Sessoli, chapter 3 in \textit{Magnetism:
Molecules to Materials}, vol. III , eds. J.S. Miller and M.
Drillon, Wiley-VCH, Weinheim (2002).}

\bibitem{tupitsyn02B}
{I. S. Tupitsyn and B. Barbara, chapter 4 in {\it Magnetism:
Molecules to Materials}, vol. III , eds. J.S. Miller and M.
Drillon, Wiley-VCH, Weinheim (2002).}

\bibitem{luis02B}
{F. Luis, F. L. Mettes, and L. J. de Jongh, chapter 5 in {\it
Magnetism: Molecules to Materials}, vol. III , eds. J.S. Miller
and M. Drillon, Wiley-VCH, Weinheim (2002).}


\bibitem{sessoli93N}
{R. Sessoli, D. Gatteschi, A. Caneschi, and M. A. Novak, Nature
(London) \textbf{365}, 141 (1993).}

\bibitem{gatteschi94S}
{D. Gatteschi, A. Caneschi, L. Pardi, and R. Sessoli, Science
\textbf{265}, 1054 (1994).}

\bibitem{friedman96PRL}
{J. R. Friedman, M. P. Sarachik, J. Tejada, and R. Ziolo, Phys.
Rev. Lett. \textbf{76}, 3830 (1996).}

\bibitem{Hernandez96}
{J. M. Hern\'{a}ndez, X. X. Zhang, F. Luis, J. Bartolom\'{e}, J.
Tejada, and R. Ziolo, Europhys. Lett. \textbf{35}, 301 (1996)}.

\bibitem{thomas96N}
{L. Thomas, F. Lionti, R. Ballou, D. Gatteschi, R. Sessoli, B.
Barbara, Nature (London) \textbf{383}, 145 (1996).}

\bibitem{sangregorio97PRL}
{C. Sangregorio, T. Ohm, C. Paulsen, R. Sessoli, and D. Gatteschi,
Phys. Rev. Lett. \textbf{78}, 4645 (1997).}

\bibitem{aubin98JACS}
{S. M. Aubin, N. R. Dilley, L. Pardi, J. Krzystek, M. W. Wemple,
L.-C. Brunel, M. B. Maple, G. Christou, and D. N. Hendrickson, J.
Am. Chem. Soc. \textbf{120}, 4991 (1998).}

\bibitem{garg93EPL}
{A. Garg, Europhys. Lett. \textbf{22}, 205 (1993).}


\bibitem{wernsdorfer99S}
{W. Wernsdordfer and R. Sessoli, Science \textbf{284}, 133
(1999).}

\bibitem{mettes01PRB}
{F. L. Mettes, F. Luis, and L. J. de Jongh, Phys. Rev. B
\textbf{64}, 174411 (2001).}

\bibitem{luis00PRL}
{F. Luis, F. L. Mettes, J. Tejada, D. Gatteschi, and L. J. de
Jongh, Phys. Rev. Lett. \textbf{85}, 4377 (2000).}



\bibitem{fernandez00PRB}
{J. F. Fern\'andez and J. J. Alonso, Phys. Rev. B \textbf{62}, 53
(2000); \textit{ibid.} \textbf{65}, 189901(E) (2002).}

\bibitem{martinez01EPL}
{X. Mart\'{i}nez-Hidalgo, E. M. Chudnovsky and A. Aharony,
Europhys. Lett.
  \textbf{55}, 273 (2001).}

\bibitem{white93PRL}
{S. J. White, M. R. Roser, J. Xu, J. T. van der Noordaa, and L. R.
Corruccini, Phys. Rev. Lett. \textbf{71}, 3553 (1993).}

\bibitem{roser90PRL}
{M. R. Roser and L. R. Corruccini, Phys. Rev. Lett. \textbf{65},
1064 (1990).}

\bibitem{noteorder}
{Although one might have thought the ordering of nuclear moments
to present an example, this certainly does not hold for the simple
metals (Cu, Ag), for which it turns out that exchange interactions
of the RKKY type play an important if not overwhelming role. A
similar remark holds for many rare-earth compounds.}

\bibitem{lis80AC}
{T. Lis, Acta Cryst. \textbf{B36}, 2042 (1980).}

\bibitem{wieghardt84AC}
{K. Wieghardt, K. Pohl, I. Jibril, and G. Huttner, Angew. Chem.
Int. Ed. Engl. {\bf 23}, 77 (1984).}

\bibitem{aubin96JACS}
{S. M. Aubin, M. W. Wemple, D. M. Adams, H.-L. Tsai, G. Christou,
and D. N. Hendrickson, J. Am. Chem. Soc. {\bf 118}, 7746 (1996)}.

\bibitem{aliaga01POLY}
{N. Aliaga, K. Folting, D. N. Hendrickson and G. Christou,
Polyhedron \textbf{20}, 1273 (2001).}

\bibitem{mettes01POLY}
{F. L. Mettes, G. Arom\'{\i}, F. Luis, M. Evangelisti, G.
Christou, D. Hendrickson, and L. J. de Jongh, Polyhedron
\textbf{20}, 1459 (2001).}

\bibitem{fernandez02PRB}
{J. F. Fern\'andez, Phys. Rev. B \textbf{66}, 064423 (2002).}

\bibitem{evangelisti04PRL}
{M. Evangelisti, F. Luis, F. L. Mettes, N. Aliaga, G. Arom\'{i},
J. J. Alonso, G. Christou, and L. J. de Jongh, Phys. Rev. Lett.
\textbf{93}, 117202 (2004).}

\bibitem{morello03PRL}
{A. Morello, F. L. Mettes, F. Luis, J. F. Fern\'{a}ndez, J.
Krzystek, G. Arom\'{\i}, G. Christou, and L. J. de Jongh, Phys.
Rev. Lett. \textbf{90}, 017206 (2003)}.

\bibitem{aromi99JACS}
{G. Arom\'{\i}, M. J. Knapp. J.-P. Claude, J. C. Huffman, D. N.
Hendrickson, and G. Christou, J. Am. Chem. Soc. \textbf{121}, 5489
(1999).}

\bibitem{affronte02PRB}
{M. Affronte, J. C. Lasjaunias, W. Wernsdorfer, R. Sessoli, D.
Gatteschi, S. L. Heath, A. Fort, and A. Rettori, Phys. Rev. B
\textbf{66}, 064408 (2002).}

\bibitem{yamaguchi02JPSJ}
{A. Yamaguchi, N. Kusumi, H. Ishimoto, H. Mitamura, T. Goto, N.
Mori, M. Nakano, K. Awaga, J. Yoo, D. N. Hendrickson, and G.
Christou, J. Phys. Soc. Japan \textbf{71}, 414 (2002).}

\bibitem{bino88S}
{A. Bino, D. C. Johnston, D. P. Goshorn, T. R. Halbert, and E. I.
Stiefel, Science \textbf{241}, 1479 (1988).}

\bibitem{furukawa01PRB}
{Y. Furukawa, K. Watanabe, K. Kumagai, F. Borsa, and D. Gatteschi,
Phys. Rev. B \textbf{64}, 104401 (2001).}

\bibitem{goto03PRB}
{T. Goto, T. Koshiba, T. Kubo, and K. Awaga, Phys. Rev. B
\textbf{67}, 104408 (2003).}

\bibitem{morello04PRL}
{A. Morello, O. N. Bakharev, H. B. Brom, R. Sessoli, and L. J. de
Jongh, Phys. Rev. Lett. \textbf{93}, 197202 (2004).}

\bibitem{morelloT}
{A. Morello, \textit{Quantum Spin Dynamics in Single-Molecule
Magnets}, Ph.D. Thesis, Leiden University (March 2004);
cond-mat/0404049.}

\bibitem{mettesT}
{F. L. Mettes, \textit{Quantum Phenomena in Molecular
Nanoclusters}, Ph.D. thesis, Leiden University (September 2001).}

\bibitem{benciniB}
{A. Bencini and D. Gatteschi, \textit{EPR of Exchange Coupled
Systems}, Springer-Verlag (Berlin and Heidelberg, 1990).}


\bibitem{steijger84PhyB}
{see, e.g., J. J. M. Steijger, E. Frikkee, L. J. de Jongh, and W.
J. Huiskamp, Physica \textbf{123B}, 271 (1984).}

\bibitem{mydoshB}
{J. A. Mydosh, \textit{Spin glasses: an experimental
introduction}, Taylor and Francis (London, 1993).}

\bibitem{reich90PRB} {D. H. Reich, B. Ellman, J. Yang, T. F. Rosenbaum, G. Aeppli, and D.P. Belanger, Phys. Rev. B \textbf{42}, 4631
(1990).}

\bibitem{reich87PRL} {D. H. Reich, T. F. Rosenbaum, and G. Aeppli, Phys. Rev. Lett. \textbf{59}, 1969 (1987).}

\bibitem{ghosh02S} {S. Ghosh, R. Parthasarathy, T. F. Rosenbaum, and G. Aeppli, Science \textbf{296}, 2195 (2002).}

\bibitem{roser92PRB} {M. R. Roser, J. Xu, S. J. White, and L. R. Corrucini, Phys. Rev. B \textbf{45}, 12337 (1992).}

\bibitem{cooke75JPC} {A. H. Cooke, D. A. Jones, J. F. A. Silva, and M. R. Wells, J. Phys. C: Solid State Phys. \textbf{8}, 4083 (1975).}

\bibitem{beauvillain78PRB} {P. Beauvillain, J.-P. Renard, I.
Laursen, and P. J. Walker, Phys. Rev. B \textbf{18}, 3360 (1978).}

\bibitem{als76PRL} {J. Als-Nielsen, Phys. Rev. Lett. \textbf{37},
1161 (1976).}

\bibitem{reich86PRB} {D. H. Reich, T. F. Rosenbaum, G. Aeppli, and H. J. Guggenheim, Phys. Rev. B \textbf{34}, 4956 (1986).}

\bibitem{moriya63B} {T. Moriya, \textit{Weak ferromagnetism}, chapter 3 in \textit{Magnetism}, Vol 1, Eds. G. T. Rado and H. Suhl, Academic Press, 1963.}

\bibitem{dejongh01AP}
{L. J. de Jongh and A. R. Miedema, Adv. Phys. \textbf{50}, 947
(2001).}

\bibitem{abragam70}
{A. Abragam and B. Bleaney, \textit{Electron Paramagnetic
Resonance of transition ions}, Oxford University Press (London,
1970).}

\bibitem{abragam61}
{A. Abragam, \textit{The Principles of Nuclear Magnetism}, Oxford
University Press (London, 1961).}

\bibitem{suter98JPCM}
{A. Suter, M. Mali, J. Roos and D. Brinkmann, J. Phys.: Condens.
Matter \textbf{10}, 5977 (1998).}


\bibitem{kubo02PRB} {T. Kubo, T. Goto, T. Koshiba, K. Takeda, and K. Awaga, Phys. Rev.
B \textbf{65}, 224425 (2002). See also: T. Kubo, A. Hirai and H.
Abe, J. Phys. Soc. Japan {\bf 26}, 1094, (1969).}

\bibitem{lascialfari98PRL}
{A. Lascialfari, Z. H. Jang, F. Borsa, P. Carretta, and D.
Gatteschi, Phys. Rev. Lett. {\bf 81}, 3773 (1998).}

\bibitem{salman02CM} Z. Salman, cond-mat/0209497 (2002).

\bibitem{pilawa05PRB}
{B. Pilawa, R. Boffinger, I. Keilhauer, R. Leppin, I. Odenwald, W.
Wendl, C. Berthier, and M. Horvati\'{c}, Phys. Rev. B \textbf{71},
184419 (2005).}

\bibitem{Slichter78} C. P. Slichter, {\it Principles of Magnetic Resonance},
Springer-Verlag (Berlin, 1978).

\bibitem{moriya56PTP} T. Moriya, Progr. Theor. Phys. {\bf 16}, 23 (1956).

\bibitem{moriya58JPSJ} T. Moriya and Y. Obata, J. Phys. Soc. Japan {\bf 13}, 1333 (1958).

\bibitem{moriya58JPSJb} T. Moriya, J. Phys. Soc. Japan {\bf 13}, 1344 (1958).

\bibitem{Kubo54} R. Kubo and K. Tomita, J. Phys. Soc. Japan {\bf 9}, 888, (1954).

\bibitem{Bloembergen49} N. Bloembergen, Physica (Amsterdam) {\bf 15}, 386 (1949).

\bibitem{Abragam82}  A. Abragam and M. Goldman, {\it Nuclear Magnetism:
Order and Disorder}, Clarendon press (Oxford, 1982).

\bibitem{Gunter67} T. E. Gunter and C. D. Jeffries, Phys. Rev. {\bf 159}, 290 (1967).

\bibitem{Lowe68} I. J. Lowe and D. Tse, Phys. Rev. {\bf 166}, 279 (1968).

\bibitem{leuenberger00PRB}
{M. N. Leuenberger and D. Loss, Phys. Rev. B \textbf{61}, 1286
(2000).}

\bibitem{Abragam80} A. Abragam and M. Borghini, {\it Dynamic polarization of nuclear
targets}, Chapter 8 in {\it Progress in Low Temperature Physics},
{\bf 4}, 384 (1964).

\bibitem{Khutsishvili70} G. R. Khutsishvili, {\it Diffusion and relaxation of nuclear spins
in crystals containing paramagnetic impurities}, Chapter 9 in {\it
Progress in Low Temperature Physics}, {\bf 6}, 375 (1970).

\bibitem{luis05PRL} {F. Luis, J. Campo, J. G\'{o}mez, G. J. McIntyre, J. Luz\'{o}n, and D. Ruiz-Molina, Phys. Rev.
Lett. \textbf{95}, 227202 (2005).}

\bibitem{Carlin86} R. L. Carlin, {\it Magnetochemistry}, Springer-Verlag
(Berlin, 1986). Chapter II.


\end{thebibliography}

\end{document}